# Planning Waste-to-Energy-Coupled AI Data Centers Through Grade-Matched Cooling and Corridor Screening

**Qi He [1,*], Chunyu Qu [2] and Wenjie Zuo [3]**

* Correspondence: qhe11@fordham.edu

**Abstract**

AI data-center (DC) growth is increasingly constrained by limited deliverable electricity, interconnection capacity, and cooling demand. This study develops a boundary-consistent screening framework for waste-to-energy (WtE)-coupled AI DC cooling, treating cooling as an energy service that can be supplied through grade matching rather than solely through electricity-driven mechanical chilling. The framework translates plant-side exportable heat into corridor-level planning objects by explicitly accounting for thermal attenuation, absorption-based conversion, and parasitic electricity associated with delivery and auxiliaries. Three results structure the analysis. First, a reference-case energy-service ledger shows how a representative regulated WtE plant with municipal solid-waste throughput of 1500 t/day and lower heating value of 10 MJ/kg yields ~78.1 MWth of exportable driving heat and, at a 20 km corridor, ~53.0 MWcool of delivered cooling and ~8.0 MWe of net avoided cooling electricity after parasitic debiting. Second, the coupled system is governed by operating regimes, not a single efficiency score. Under the baseline package, full thermal coverage is maintained up to ~20.9 km, the stricter quality-adjusted criterion remains positive to ~22.9 km, and the electricity–relief criterion remains positive to ~44.7 km. Third, deployment-scale translation for a 1 GW IT campus ($u = 0.70$, $L = 5$ km) implies a net grid relief of ~116.9–264.4 MW across scenario packages, while the required WtE footprint ranges from roughly three to 148 equivalent representative plants, or about 0.6–40 full-load-equivalent plants at a 25% displacement target. The contribution is a siting-ready planning framework that identifies when WtE-coupled cooling remains corridor-feasible, when it becomes hybrid and marginal, and when infrastructure scale rather than thermodynamic benefit becomes the binding constraint. It is intended as a screening tool for planning and comparison, not as a project-specific hydraulic or plant-cycle design.

**Keywords**: sustainable thermal management; waste-to-energy; absorption cooling; thermal energy storage; renewable energy conversion; AI data centers; thermoeconomics; grid-capacity relief

## 1. Introduction

AI data center (DC) expansion is increasingly constrained not by model architectures alone, but by the joint availability of deliverable electricity, interconnection capacity, and heat rejection at scale. Recent assessments show that DCs already account for a material electricity load and that AI-driven growth could raise demand sharply over the next decade [1,2]. What binds in practice, however, is often local and physical rather than global and abstract. Interconnection queues, equipment lead times, and limits on thermal and water footprints increasingly determine if the compute can be deployed within a feasible electricity-cooling envelope. In this setting, the scarce object is not computed in the abstract. It is deployable compute that can be powered and cooled within infrastructure constraints.

Cooling sits at the center of that envelope because it affects both operating cost and feasible power density. Conventional mechanical cooling uses high-grade electricity to move low-grade heat to ambient. Even with containment, economizers, and improved controls, ultimate heat rejection still depends on compressor and fan work that scales with thermal load. This has accelerated the move toward liquid-assisted architectures. Yet the dominant forms of liquid cooling mainly improve heat transfer near the chip or rack, while many facilities still rely, partially or fully, on electricity-driven chillers, cooling towers, dry coolers, or hybrid systems for heat rejection [3]. Cooling therefore remains a first-order determinant of marginal electricity demand even as thermal density improves.

A different pathway follows from grade matching. Cooling is an energy service, and it does not require high-grade electricity as its only input. Low-grade thermal energy can be converted into delivered cooling through thermally driven cycles, displacing compressor electricity at the facility boundary. Municipal solid-waste systems provide a plausible and underused source of such heat in many urban regions. Waste-to-energy (WtE) plants recover energy from municipal solid waste, and part of that thermal output can be suitable for absorption cooling. At the same time, diverting waste from landfills changes the counterfactual methane pathway. Because methane is a short-lived but high-impact greenhouse gas, the near-term climate relevance of methane-dominant baselines is well-established in science and policy [4,5]. These considerations create a system context in which waste infrastructure and AI infrastructure can be evaluated as coupled services.

He and Qu (2025) provided a narrative entry point for waste-sector interventions motivated by AI-era grid resilience [6]. The present paper is narrower and more decision-oriented. It studies the coupling mechanism that converts WtE thermal output into delivered cooling for AI DCs, and the conditions under which that coupling remains feasible once spatial deliverability and parasitic electricity are treated explicitly. A second issue is practical credibility. WtE is often viewed as intrinsically dirty, a perception that is historically understandable but technically incomplete. Modern plants operate under stringent pollutant control regimes, and leading jurisdictions encode these requirements in enforceable standards in the European Union, the United States, and China [7–9]. Large-sample empirical evidence likewise treats pollutant control and resource recovery as measurable performance dimensions rather than fixed constraints [10]. The analysis therefore conditions on a modern regulated plant and focuses on the following question: when can residual WtE heat support DC cooling under realistic corridor constraints?

To answer that question, the paper develops a screening-level infrastructure model that links four objects: a representative WtE exportable thermal stream, a heat-delivery corridor with distance-dependent attenuation, absorption-based conversion of heat into cooling, and an electricity accounting identity that translates delivered cooling into an avoided mechanical chilling net of parasitic loads. The coupled system is treated as a boundary-consistent input–output structure rather than as a project-specific hydraulic design. This makes it possible to express the main planning objects in a small set of interpretable quantities: an energy-service ledger, a heat-driven coverage ratio, a quality-adjusted corridor criterion, and scenario-specific feasibility bands. The paper is therefore not intended as a component-resolved thermodynamic design or plant-internal exergy-destruction study, nor does it claim universal scalability of WtE-coupled AI DC cooling. Its objective is narrower and more practical: to identify when such coupling remains infrastructure-feasible within corridor, coverage, auxiliary-load, and feedstock-quality bounds before project-specific design begins.

Three findings organize the analysis. First, the reference-case energy-service ledger converts exportable WtE heat into delivered cooling and then into gross and net avoided cooling electricity under explicit attenuation, conversion, and parasitic debiting. In the baseline reference case, a representative regulated WtE plant provides about 78.1 MWth of exportable driving heat. At a 20 km corridor, this translates into 70.7 MWth of delivered driving heat, 53.0 MWcool of delivered cooling, 15.1 MWe of gross avoided cooling electricity, and 8.0 MWe of net avoided cooling electricity after parasitics. Second, the coupled system is governed by operating regimes rather than by a single efficiency score. Under the baseline parameterization, full thermal coverage is maintained up to about 20.9 km, while the stricter quality-adjusted threshold remains positive to about 22.9 km; by contrast, the electricity–relief criterion remains positive to about 44.7 km. These thresholds distinguish full coverage, partial but viable operation, and non-viable corridor conditions. Third, deployment-scale translation separates benefit from implementability. For a 1 GW IT campus at $u = 0.70$ and $L = 5$ km, net grid relief ranges from about 116.9 MW to 264.4 MW across scenario packages, while the implied WtE footprint ranges from roughly three to 148 equivalent representative plants under full-coverage sizing, or about 0.6 to 40 full-load-equivalent plants at a 25% displacement target. The implication is straightforward: large grid relief does not automatically imply easy deployment.

Two interfaces complete the decision framing. The first is a reporting boundary for sustainability and ESG interpretation, built around auditable waste diversion volumes, explicit landfill counterfactuals, and grid-electricity displacement reported conditional on coverage regime and corridor distance. The second is an economic interface, introduced through a levelized cost of cooling-compute service (LCOC) under a common accounting boundary. These additions do not replace the physical screening logic. They translate it into planning-relevant reporting and cost objects. Taken together, the contribution of the paper is a conservative but decision-ready framework for screening WtE-AIDC pairings, clarifying when grade-matched cooling can relieve cooling-related electricity demand, when corridor losses and auxiliary burdens dominate, and when deployment-scale requirements become the binding constraint.

The present manuscript is not intended as a component-resolved thermodynamic design or plant-internal exergy-destruction study. It is positioned as a boundary-consistent screening framework that translates plant-side exportable heat into corridor-relevant cooling, electricity relief, and quality-adjusted feasibility objects for WtE–AIDC pairing. This study also does not claim the universal scalability of WtE-coupled AI DC cooling. Its objective is narrower and more actionable: to identify whether such coupling remains infrastructure-feasible within corridor, coverage, auxiliary-load, and feedstock-quality bounds before project-specific design studies are undertaken.

### *1.1. Related Work and Positioning*

Recent work on DC cooling has largely focused on reducing support energy through facility design, controls, and alternative cooling architectures. In practice, this literature is often organized around facility-level metrics such as PUE, which remains useful as a standardized accounting ratio between IT load and total facility energy use [9]. Yet even with the post-2020 shift toward liquid cooling, recent reviews emphasize that many deployments still rely on electricity-driven rejection systems at the plant boundary, including chillers, dry coolers, towers, or hybrid arrangements [2]. In this sense, much of the existing DC literature remains centered on improving an electric-to-cooling pathway rather than replacing part of that pathway with an external thermal service. Prior studies are adjacent in method or application, but they do not directly formulate the corridor-level planning object developed here.

A second literature examines waste-heat recovery, district-energy integration, and DC-integrated energy systems. Reviews synthesize recovery pathways, utilization modes, and enabling conditions such as temperature level, temporal matching, storage, and network coordination [10–12]. Related studies on integrated energy systems, flexible dispatch, and coordinated scheduling further show that DCs can be modeled within broader multi-energy infrastructures [13-17]. These studies are highly relevant, but they usually optimize operation within a given integrated system, or evaluate heat recovery once a usable thermal link is already presumed. They do not typically formulate the specific siting problem considered here: how exportable thermal energy from a regulated WtE plant translates, under corridor attenuation and auxiliary burdens, into a planning-relevant cooling service for an AI DC.

A third body of work addresses district heating and cooling networks, including long-distance thermal delivery, storage, and absorption-based conversion. These studies make clear that feasibility depends not only on device performance but also on network configuration, pumping work, temperature levels, and spatial separation [16-22]. They therefore point toward an infrastructure reality that is central to the present paper: a heat source is not automatically usable at the point of demand. However, most of this literature remains oriented toward district-network design, staged heat-pump systems, or equipment-level thermodynamic performance, rather than a corridor-screening framework that converts plant-side exportable heat into delivered cooling, avoided electricity, and explicit feasibility thresholds for WtE–AIDC pairing.

The WtE literature provides the supply-side counterpart. Recent work emphasizes heterogeneity in plant configuration, emissions control, thermal recovery, infrastructure planning, and allocation rules within district-energy settings [23-27]. This is important as the usable residual thermal stream is shaped not only by plant technology but by network interface and system boundary choices. The logic motivates the boundary here: conditional on a modern regulated WtE facility, the relevant planning question is not plant-internal optimization alone, but whether residual thermal output can displace high-grade electricity otherwise used for cooling under explicit corridor and auxiliary constraints.

Thermoeconomic and exergy methods remain useful in this setting, not because the present study aims to deliver a component-resolved exergy design, but because grade matching requires a quality-consistent accounting framework [28,29]. Here, exergy is used as a boundary-consistent screening criterion for comparing an electricity-driven cooling baseline with a coupled thermal-cooling pathway. The contribution of this paper therefore lies not in rederiving standard thermodynamic principles, but in synthesizing adjacent studies into a planning-oriented screening framework for WtE-coupled AI DC cooling. That positioning is made explicit in the revised Sections 3.1–3.5, where the analysis is organized as an energy-service ledger, an operating-regime map, a quality-adjusted corridor criterion, a feedstock-distance sensitivity surface, and scenario feasibility bands.

### *1.2. Gap and Contribution*

This paper is among the first to formulate WtE-coupled AI DC cooling as a boundary-consistent corridor-screening problem rather than as a device-level cooling study, plant-internal thermodynamic design exercise, or generic waste-heat integration concept. The originality of this paper lies in formulating WtE-coupled AI data-center cooling as a boundary-consistent corridor-screening and planning problem, rather than as a device-level cooling study or a plant-internal thermodynamic design exercise. The contribution is the construction of a unified object chain from exportable heat to delivered cooling, net electricity relief, regime thresholds, and deployment-scale infrastructure requirements.

The contribution is fourfold. First, the paper develops a reference-case energy-service ledger that converts exportable WtE heat into delivered driving heat, delivered cooling, gross avoided cooling electricity, parasitic debit, and net avoided cooling electricity. This makes the coupled system auditable at the screening stage and avoids collapsing the problem into a single opaque efficiency number. Second, it introduces a regime-based corridor framework that distinguishes full coverage, partial but viable operation, and non-viable corridor conditions, and then tightens that logic through a quality-adjusted criterion that credits delivered cooling by thermodynamic quality rather than by thermal quantity alone. Third, it converts those continuous results into scenario-specific feasibility bands that are interpretable for siting and planning under uncertainty. Fourth, it extends the screening objects to deployment scale, separating energy benefit from implementability for a 1 GW IT campus, while linking the thermodynamic results to decision interfaces for LCOC and boundary-disciplined ESG reporting. The result is not a universal deployment claim or a project-specific engineering design. It is a compact planning framework that identifies when WtE–AIDC coupling remains plausible, when it becomes marginal, and when corridor losses, auxiliary burdens, and infrastructure scale become the binding constraints.

## 2. Methodology and Data—Integrated System and Thermoeconomic Model

This section introduces the integrated WtE-AIDC concept using an input–output accounting lens, then formalizes a thermoeconomic model for the coupled WtE-AI DC architecture. We intentionally abstract from equipment-level engineering (pipe routing, welding, chiller internals, turbine configuration) and focus on the minimum set of energy-service flows that determine feasibility and economic value: waste-derived primary energy, electricity, recoverable heat, delivered cooling service, grid purchases, and parasitic requirements. Accordingly, the present model does not endogenously solve a plant-specific turbine cycle or a routed thermo-hydraulic corridor. Instead, plant-side heat availability is represented through a net exportable thermal parameter, while corridor-side hydraulic burdens are represented through reduced-form decay and pumping terms anchored to published engineering ranges.

Figure 1 defines the accounting topology of the integrated WtE–AIDC system. Within the integrated boundary, the WtE plant converts waste feedstock $(W, LHV)$ into two usable outputs: electricity and net exportable heat. WtE electricity can offset purchased grid electricity, while net exportable heat enters the coupling link and is converted into delivered cooling for the AIDC campus. At the screening level, the coupling link is represented by three quantities only: the thermal-delivery factor $\eta_{tr}(L)$, the absorption cooling performance $COP_{abs}$, and the parasitic electricity burden $W_{par}(L)$. On the demand side, the AIDC campus is characterized by IT load $P_{IT}$, cooling requirement $Q_{req}$, and compute output. Grid electricity $E_{grid}$ supplies the IT load, fixed overhead, and any residual cooling not met by the heat-driven pathway. The dashed branch denotes the standalone benchmark, in which cooling is provided entirely by mechanical chilling with $COP_m$.

This topology is not a detailed engineering layout. It is the minimum accounting structure needed to compare coupled and standalone operation on a common boundary. Its purpose is to establish, before any project-specific design, how waste-derived energy is translated into cooling service, avoided grid electricity, and residual mechanical demand. Distance enters the model only through two corridor effects, thermal deliverability $\eta_{tr}(L)$ and parasitic burden $W_{par}(L)$. That choice keeps the framework compatible with a wide range of corridor-scale configurations while preserving transparent bookkeeping. Figure 1 therefore serves as the organizing model for Section 2: it fixes the system boundary, identifies the relevant energy-service flows, and defines the objects that later appear in the thermoeconomic screening.

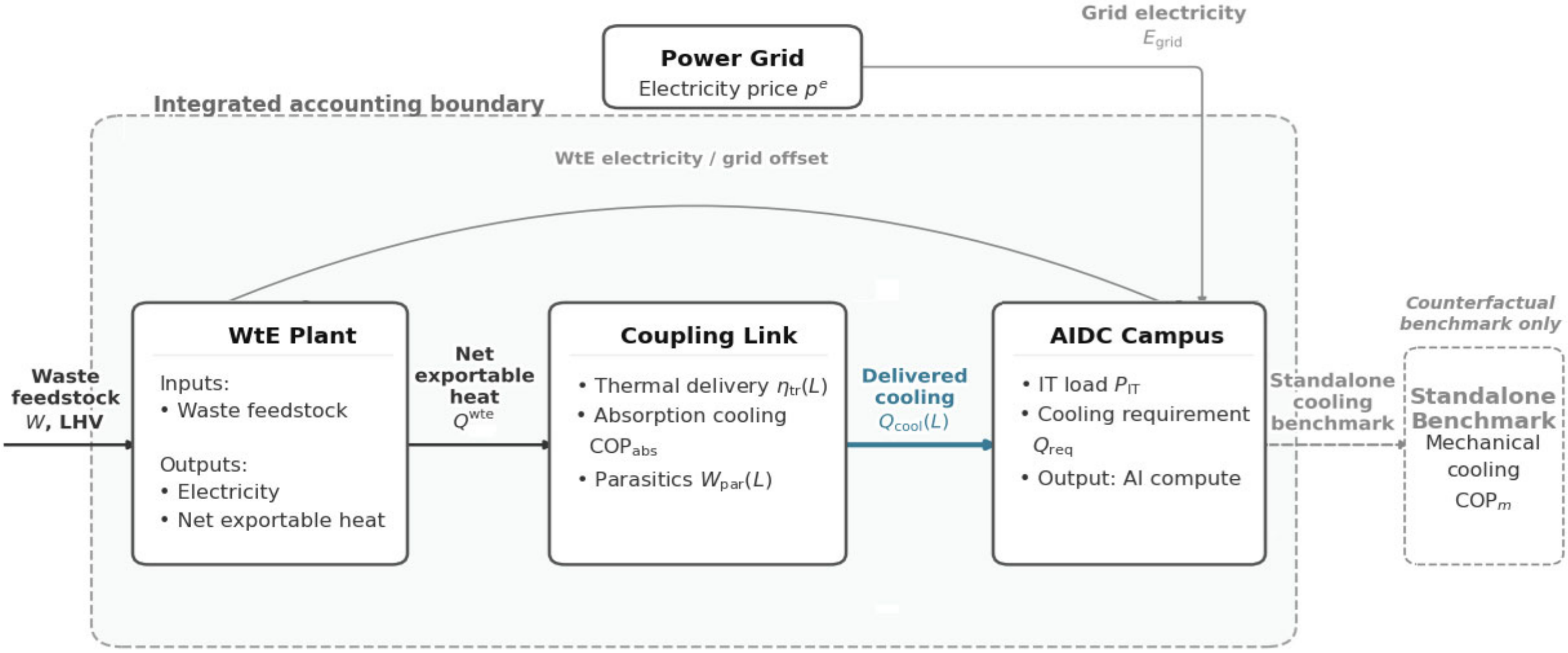

**Figure 1.** Integrated-system accounting topology and standalone benchmark. Notes: The figure defines the integrated accounting boundary linking the WtE plant, the coupling link, and the AIDC campus. Blue labels indicate the coupled cooling-service pathway, whereas gray dashed elements denote the standalone benchmark/counterfactual only. Within this boundary, waste feedstock is converted into electricity and net exportable heat; thermal service is then transmitted through the coupling link, where distance-dependent deliverability $\eta_{tr}(L)$, absorption cooling $COP_{abs}$, and parasitic electricity $W_{par}(L)$ enter the accounting. Delivered cooling, $Q_{cool}(L)$, offsets campus cooling demand, while WtE electricity may offset pur-

chased grid electricity. The dashed box and dashed arrow on the right denote the counterfactual standalone mechanical-cooling benchmark only and are not part of the integrated physical system. Equipment-level internals are intentionally abstracted to preserve system-level accounting focus.

We model all flows as steady-state rates based on former work and setting [6]. A dot over a variable denotes a rate (e.g., MW, MJ/s). The integrated system boundary includes (i) the WtE plant converting waste chemical energy into electricity and recoverable heat, (ii) a thermal bridge delivering usable heat to the DC, and (iii) the conversion of heat to cooling via absorption chilling. Key notations are listed in Table 1.

**Table 1.** List of symbols and thermodynamic parameters.

| Symbol | Definition | Units |
|---|---|---|
| $\dot{m}_w$ | Waste mass flow rate | kg/s |
| $LHV$ | Lower heating value of waste | MJ/kg |
| $\dot{E}_{in}$ | Primary energy input from waste $(\dot{m}_w \cdot LHV)$ | MW |
| $\eta_c$ | Effective combustion/boiler utilization efficiency | - |
| $\eta_e$ | Net electrical conversion efficiency (WtE) | - |
| $\alpha_h$ | Recoverable heat fraction of non-electric output (heat recovery factor) | - |
| $\dot{W}_e$ | Net electric output from WtE | MW |
| $\dot{Q}_h$ | Net exportable heat available for external cooling service | MW |
| $L$ | Separation distance (WtE to DC)/thermal bridge length | km (or m) |
| $\eta_{tr}(L)$ | Effective thermal delivery factor (reduced-form) | - |
| $\beta$ | Thermal decay coefficient in $\eta_{tr}(L)$ | 1/km |
| $COP_{abs}$ | Absorption chiller coefficient of performance | - |
| $COP_m$ | Mechanical (electric) chiller COP | - |
| $\dot{Q}_{cool}$ | Cooling rate supplied to DC | MW |
| $\dot{W}_{IT}$ | DC IT electrical load | MW |
| $\gamma$ | Cooling requirement per IT load $\left(\dot{Q}_{req} = \gamma\dot{W}_{IT}\right)$ | - |
| $\dot{W}_{aux}$ | Non-cooling auxiliary DC load (fans, UPS losses, lighting) | MW |
| $\dot{W}_{par}$ | Parasitic work for thermal delivery and absorption (pumps, controls) | MW |
| $T_0$ | Ambient reference temperature | K |
| $T_c$ | Chilled-water/refrigeration temperature | K |
| $\phi_c$ | Cooling exergy factor $= \frac{T_0}{T_c} - 1$ | - |

*2.1. System Topology Design and Boundaries*

Figure 1 represents the integrated system as three facility nodes connected by energy-service pathways. The WtE plant converts municipal solid waste $W$ with heating value LHV into electricity $E^{wte}$ and net exportable heat $Q^{wte}$. Electricity can be supplied to the AIDC campus through an on-site connection or, equivalently, can offset grid purchases. Recoverable heat enters a coupling link, modeled here as a black box that converts heat into cooling through absorption chilling. For campus-scale accounting, this link is characterized by three quantities: the absorption performance parameter $COP_{abs}$, the distance-dependent thermal delivery factor $\eta_{tr}(L)$, and the parasitic electricity demand $W_{par}(L)$, which captures pumping and auxiliary loads associated with thermal transport and conversion. The AIDC campus consumes IT power $P_{IT} = uP_{IT}^{max}$ and requires cooling $Q_{req} = \gamma P_{IT}$ to deliver compute service K. Grid electricity enters as $E^{grid}$ at price $p^e$, supplying the IT load, fixed overhead, and any residual cooling not covered by the heat-driven pathway. The dashed connection in Figure 1 denotes the standalone counterfactual. Without coupling, the campus must meet its cooling requirement through mechanical chilling with the coefficient of performance $COP_m$, implying a cooling electricity requirement proportional to $Q_{req}/COP_m$. In the integrated system, the heat-driven pathway partially or fully displaces this electricity component, creating the physical basis for lower purchased electricity and improved operating energy metrics.

Two boundary clarifications matter. First, Figure 1 does not specify a unique engineering layout. It defines the accounting objects needed to compare coupled and standalone operation. Second, distance L enters only through deliverability and parasitic burden, namely $\eta_{tr}(L)$ and $W_{par}(L)$. The topology is therefore compatible with a broad set of corridor-scale co-location designs. This is consistent with the paper's objective: to establish the viability and value logic of integrated WtE-AIDC deployments without committing to a single hardware blueprint.

*2.2. Assumptions and Accounting Rules*

Figure 2 gives the operational interpretation of the coupled system under variability. The thermal stream entering the cooling pathway is not the plant's total recoverable energy, but the net exportable heat available to cooling after plant-side commitments. The present paper does not solve an hourly dispatch problem between electricity generation for computation and thermal drive for absorption cooling. Instead, it treats the plant-side split at screening level through the exportable-heat representation and focuses on the campus-side cooling-service consequences of that net available thermal stream. On the campus side, the intended interpretation is a reliability-first hybrid operation: IT service and the cooling setpoint are preserved first, WtE-driven absorption cooling is used whenever thermal service is available, and the mechanical-chilling path absorbs residual and peak demand. Accordingly, the framework should be read as a cooling-electricity displacement model with explicit fallback support, rather than as a claim that temperature-critical service is supplied exclusively by the WtE-coupled stream.

(**a**)

Coupled System Boundary
Thermal Corridor
Net Exportable Heat
WtE Plant
MSW Feedstock
to cooling after plant-side commitments
Peak cooling demand
delivery decay: $e^{-\beta d}$
Single-Effect Absorption Chiller
Delivered Cooling
AIDC Cooling Header
AI Data Center Thermal Load
Backup / Residual Cooling
Grid Electricity
Mechanical Chiller / Backup Cooling
Campus cooling priority
• Maintain IT service & setpoint
• Maximize WtE cooling
• Mechanical covers residual/peak

(**b**)

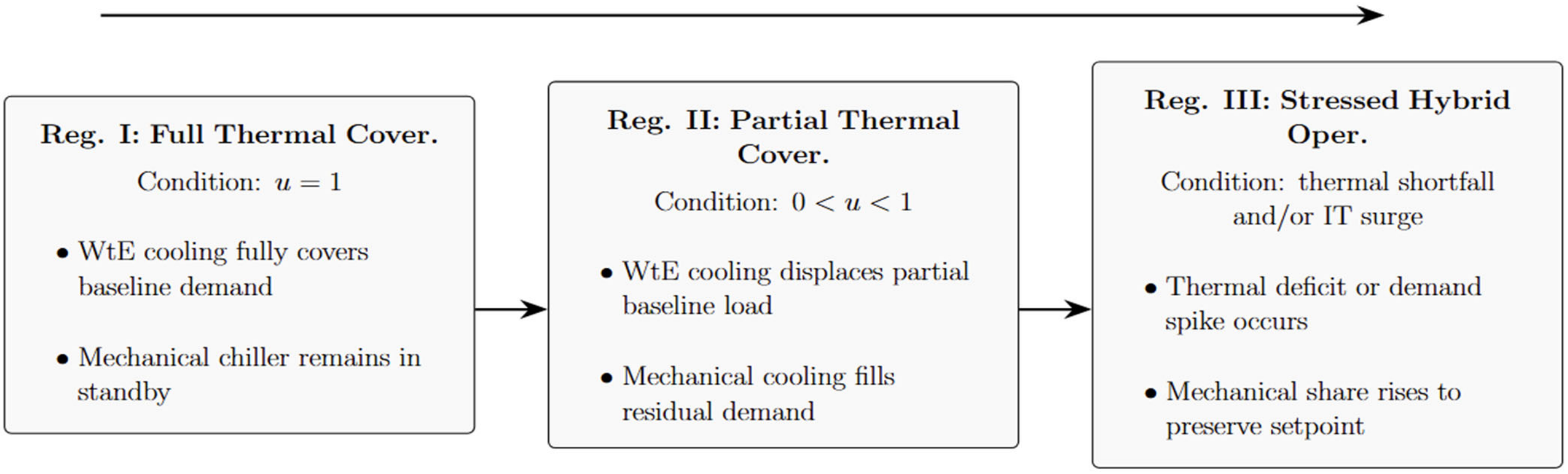

**Figure 2.** Hybrid cooling architecture and operating regimes under net exportable heat and peak cooling demand. Notes: (a) Hybrid cooling architecture of the coupled WtE–AIDC system. Net exportable heat from the WtE plant is delivered through the thermal corridor to a single-effect absorption chiller and then to the AIDC cooling header. Solid arrows denote the primary thermal/cooling-service pathway and direct demand linkages, whereas dashed arrows and dashed boxes denote backup/residual cooling paths, contextual links, or boundary elements rather than the main coupled pathway. (b) Operating-regime map under increasing stress from lower net available heat and/or higher peak cooling demand. Regime I represents full thermal coverage, Regime II partial thermal coverage, and Regime III stressed hybrid operation in which the mechanical-cooling share rises to preserve the cooling setpoint.

*2.3. Minimal Energy-Flow Formulation*

(1) WtE conversion

The primary energy input rate from waste is

$$\dot{E}_{in} = \dot{m}_w \cdot LHV \tag{1}$$

Net electricity output is modeled as

$$\dot{W}_e = \eta_e \, \eta_c \, \dot{E}_{in} \tag{2}$$

Recoverable thermal output (for downstream utilization):

$$\dot{Q}_h = \alpha_h \left( \eta_c \dot{E}_{in} - \dot{W}_e \right) \tag{3}$$

Equation (3) treats heat recovery as a plant-level recoverable fraction of the non-electric useful energy inside the boundary.

(2) Thermal delivery (spatial feasibility constraint)

Thermal delivery to the DC is constrained by distance L through a reduced-form delivery factor:

$$\dot{Q}_{del} = \eta_{tr}(L)\,\dot{Q}_h, \eta_{tr}(L) = exp(-\beta L) \tag{4}$$

This $\eta_{tr}(L)$ term compactly captures aggregate thermal losses, imperfect insulation, and operational constraints (including pumping/controls) without claiming detailed pipe-network thermo-hydraulics. In the present screening framework, the thermal-delivery term is interpreted as an aggregate corridor loss parameter rather than as a buried-pipe heat-transfer simulation. The baseline assumption is modern pre-insulated district-energy piping, and the decay band is anchored to published linear heat-loss ranges reported in W/m and then translated into an implied fractional loss per kilometer using a representative trunk thermal flow. Seasonal ground-temperature and ambient-condition variation are not modeled as a separate transient soil-heat-transfer problem; instead, their effect is conservatively absorbed into the scenario band for the corridor loss parameter together with insulation quality and the effective temperature differential. It is therefore appropriate for comparative statics and screening-level feasibility.

To make the screening expression more explicit, we distinguish plant-side and corridor-side engineering proxies. On the plant side, exportable driving heat is represented as

$$\dot{Q}_{\exp} = \phi_{\text{th,net}}\dot{m}_w LHV \tag{5}$$

where $\phi_{\text{th,net}}$ is a net exportable thermal parameter intended to bracket internal heat use, boiler/cycle performance, and plant-side heat-extraction choices at screening level. In the notation of the present model, this exportable driving heat is the recoverable-heat term introduced in Equation (3), $\dot{Q}_{\exp} \equiv \dot{Q}_h$ (in the later sections we stick to $\dot{Q}_{\exp}$ for all equivalent expression of $\dot{Q}_h$). The model therefore does not infer a unique power-to-heat ratio from a specified turbine configuration such as back-pressure or extraction-condensing operation. Instead, such engineering differences are absorbed into the net exportable thermal parameter, which is interpreted as the heat stream available for external cooling service after internal uses, plant-cycle configuration, and extraction priorities are accounted for. This representation is intended for screening and corridor-planning analysis rather than plant-specific turbine-cycle design, and it can be replaced by a project-level heat balance once site data become available.

On the corridor side, hydraulic burdens are likewise represented in reduced form rather than through a routed thermo-hydraulic network solution. Specifically, the parasitic electricity term introduced below is decomposed as $\dot{W}_{\text{par}} = \dot{W}_{\text{pump}} + \dot{W}_{\text{aux,abs}}$, with corridor pumping represented as $\dot{W}_{\text{pump}} = I_{\text{pump}}\dot{Q}_{\text{del}}$, where $I_{\text{pump}}$ is a reduced-form hydraulic-burden proxy, $\dot{W}_{\text{pump}}$ denotes corridor pumping and recirculation electricity associated with thermal delivery, and $\dot{W}_{\text{aux,abs}}$ denotes absorber-plant auxiliaries, including cooling-tower fan/pump equivalents and controls. In this sense, route-specific pressure-drop effects are not solved explicitly from hydraulic network equations; rather, their impact is absorbed into the distance-delivery term in Equation (4) together with the parasitic-load formulation below, which jointly bracket plausible routing, diameter, temperature-level, and hydraulic-station conditions at screening level. This representation preserves an auditable system-level boundary while avoiding false precision at the pre-design stage.

(3) Absorption cooling transformation

Delivered heat is converted into cooling via absorption chilling:

$$\dot{Q}_{cool} = COP_{abs}\,\dot{Q}_{del} \tag{6}$$

The DC's cooling requirement is parameterized as $\dot{Q}_{req} = \gamma \dot{W}_{IT}$. Define the heat-driven coverage share as $f = min\left\{1, \frac{\dot{Q}_{cool}}{\dot{Q}_{req}}\right\}$. The residual cooling demand $(1-f)\dot{Q}_{req}$ is met by mechanical chilling, requiring electricity:

$$\dot{W}_{cool,m} = \frac{(1-f)\dot{Q}_{req}}{COP_m} \tag{7}$$

Residual mechanically supplied cooling should also be interpreted as the reliability-preserving fallback path under transient reductions in delivered thermal cooling. The model therefore does not assume that server-grade chilled-water temperature control relies exclusively on the WtE-coupled stream. Finally, parasitic work $\dot{W}_{par}$ is modeled as a transparent reduced-form load that includes two screening-level components: (i) corridor pumping and recirculation electricity associated with long-distance thermal delivery, and (ii) absorber-plant auxiliaries, including cooling-tower fan/pump equivalents and controls. This representation is intended to prevent "free-cooling" artifacts while preserving an auditable system-level accounting boundary.

$$\dot{W}_{par} = \kappa_0 + \kappa_1 \dot{W}_{IT} + \kappa_2 L \tag{8}$$

where $\kappa_0, \kappa_1, \kappa_2$ can be scenario-calibrated. This keeps the model auditable while preserving the focus on system accounting.

*2.4. Metrics and Decision Criteria*

(1) Extended PUE—auditable under heat-driven cooling

Conventional PUE is electricity-only and becomes non-comparable once cooling is supplied thermally. We therefore define two complementary metrics:

(i) Electric PUE (facility electricity only):

$$PUE_{elec} \equiv \frac{\dot{W}_{IT} + \dot{W}_{aux} + \dot{W}_{cool,m} + \dot{W}_{par}}{\dot{W}_{IT}} \tag{9a}$$

This metric answers how much grid electricity the DC draws per unit IT load in coupling.

(ii) $PUE_{sys}$ (within the coupled boundary) is a system-boundary accounting ratio that tracks the total energy burden required to deliver computing and cooling services relative to IT load. In this sense, it is conceptually close to an Energy-Service Intensity (ESI). We retain the notation $PUE_{sys}$ for consistency with the paper's derivations, but it is not an ISO/IEC PUE metric and is not intended for cross-study benchmarking. Its role is purely internal: to maintain conservation-consistent accounting when part of the cooling service is supplied by non-electric thermal energy.

$$PUE_{sys} \equiv \frac{\dot{W}_{IT} + \dot{W}_{aux} + \dot{W}_{cool,m} + \dot{W}_{par} + \dot{Q}_{del}}{\dot{W}_{IT}} \tag{9b}$$

Here $\dot{Q}_{del}$ is counted as an energy input supporting computing, ensuring comparability across electricity-driven and heat-driven cooling. This avoids the critique that PUE improves simply because energy is shifted from electricity to heat. $PUE_{elec}$ captures grid capacity relief and electric OpEx; $PUE_{sys}$ captures total energy intensity within the integrated system boundary.

(2) Systemic Exergy Efficiency ($\eta_{ex}$)

Energy quantities alone do not distinguish high-grade electricity from low-grade heat. We therefore define a systemic exergy efficiency, treating waste chemical energy as the primary exergy input (a standard screening approximation for comparative statics):

$$Ex_{in} \equiv \eta_c \dot{E}_{in} \tag{10}$$

Useful exergy output includes the electricity and the exergy of delivered cooling. In the present paper, quality-adjusted accounting refers specifically to crediting delivered cooling not by its full thermal quantity, but by its exergy relative to ambient conditions. The exergy of a cooling effect at temperature $T_c$ relative to ambient $T_0$ is

$$\phi_c \equiv \frac{T_0}{T_c} - 1, Ex_{cool} \equiv \phi_c \dot{Q}_{cool} \tag{11}$$

Thus total useful exergy output is

$$Ex_{out} \equiv \dot{W}_e + Ex_{cool} \tag{12}$$

Systemic exergy efficiency:

$$\eta_{ex} \equiv \frac{Ex_{out}}{Ex_{in}} = \frac{\dot{W}_e + \phi_c \dot{Q}_{cool}}{\eta_c \dot{E}_{in}} \tag{13}$$

$\eta_{ex}$ is a quality-adjusted utilization index, which increases when low-grade heat is converted into a service (cooling) that offsets high-value electricity use, i.e., when the system reduces exergy destruction through better grade matching. Because the delivered cooling term satisfies $\dot{Q}_{cool} = COP_{abs}\dot{Q}_{del}$, a lower absorption COP directly reduces the credited cooling service before the exergy conversion is applied. By contrast, the advantage of a modern high-$COP_m$ centrifugal chiller enters through the electric-displacement benchmark, where the same cooling service would require only $\dot{W}_{cool,m} = \dot{Q}_{cool}/COP_m$ units of electricity. Contrastly, the advantage of a modern high-$COP_m$ centrifugal chiller enters through the electric-displacement benchmark where the same cooling service requires only $\dot{W}_{cool,m} = \dot{Q}_{cool}/COP_m$ units of electricity. Refer to Appendix A for a sufficient condition.

*2.5. Decision Metrics and Accounting Boundary*

Because the present study compares a conventional electricity-driven cooling baseline with a WtE-coupled thermal-cooling pathway, the thermodynamic screening requires a decision layer that remains comparable under a common accounting boundary. The role of this layer is not to replace the physical screening developed in Section 3, but to translate it into planning-relevant comparison objects.

In this manuscript, the levelized cost of cooling-compute service (LCOC) is introduced as a cost-normalized interface metric:

$$LCOC = \frac{\sum_t \omega_t C_t (1+r)^{-t}}{\sum_t \omega_t K_t (1+r)^{-t}}, \tag{14}$$

where $C_t$ denotes the total attributable cost net of relevant revenues in period $t$, $K_t$ is the compute-service denominator under a common accounting boundary, $r$ is the discount rate, and $\omega_t$ is an indicator or weighting term for service delivery. The purpose of LCOC is not to construct a fully parameterized project-finance model, but to provide a boundary-consistent comparison interface between the baseline configuration and the WtE-coupled alternative.

Under this boundary, the attributable cost in period t may be written as

$$C_t = CapEx_t^{ref} + CapEx_t^{couple} + OpEx_t + r_t^E E_t^{grid} + p_t^w W_t - Rev_t^{elec} \tag{15}$$

where the electricity term reflects the extent to which the cooling demand remains grid-served after thermal coupling. Using the coverage formulation developed later in Section 3, grid electricity can be expressed as

$$E_t^{grid} = E_t^{IT} + E_t^{aux} + \big(1 - f(L,u)\big)\frac{Q_{d,0}}{COP_m} + W_{par,t}(L) - E_t^{on\text{-}site} \tag{16}$$

with $f(L,u) \in [0,1]$ denoting the heat-driven coverage ratio, $Q_{d,0}$ denoting the baseline cooling requirement, $COP_m$ denoting the benchmark mechanical-cooling coefficient of performance, and $W_{par,t}(L)$ denoting the corridor-dependent parasitic electricity burden.

Sections 3.1–3.5 then supply the physical inputs required by these decision metrics. In particular, the staged energy-service ledger yields delivered cooling and net avoided electricity, the operating-regime formulation determines whether the system remains in full coverage or hybrid operation, and the quality-adjusted criterion identifies when corridor benefits remain meaningful under stricter thermodynamic screening. Section 4 interprets these methodology-defined metrics in planning and economic terms.

*2.6. Data, Parameterization, and Scenario Design*

This study is framed as a screening-level infrastructure analysis rather than a project-specific engineering design. Its inputs are taken from published, auditable ranges and assembled into a compact parameter set that can be replaced later with corridor-specific measurements. Table 2 reports the key parameters, their baseline ranges, and the scenario packages used in the main figures and sensitivity analysis (the AI specificity of the present study lies in the deployment context, including higher electricity demand growth, stronger cooling pressure, and interconnection constraints, rather than in the use of AI or machine-learning methods within the model itself).

(1) Evidence base and reproducibility boundary

Public, site-specific data on WtE–DC thermal coupling remain limited. The empirical strategy therefore relies on transparent parameterization grounded in three evidence classes: (i) DC energy accounting and oper-ating envelopes, (ii) WtE heat-balance magnitudes and exportable temperature ranges, and (iii) district-energy delivery constraints, including heat loss, pumping parasitics, and corridor-scale feasibility. Reproducibility is maintained in three ways. First, every parameter in Table 2 is reported with a unit and interpretation [26,27,30]. Second, the conservative, baseline, and aggressive cases help separate physical uncertainty from economic context. Third, reduced-form terms such as distance decay are linked back to engineering quantities reported in pipe guides and district-energy studies, so each assumption can be checked or replaced with project data [31–33].

(2) Parameter classes and data extraction

(a) The DC block is described by IT capacity (MW), utilization $u$, and a baseline electricity boundary expressed through PUE. Here PUE is used only as an accounting ratio that maps IT load into total facility electricity in the standalone benchmark, consistent with standard definitions and reporting practice [11]. Baseline ranges are drawn from sector reports and surveys on U.S. DC energy use and operating performance [2,34], together with recent assessments of AI-related system constraints such as electricity deliverability, interconnection delays, and rising cooling intensity. When source ranges differ, Table 2 reports a midpoint baseline and Table 3 carries the uncertainty through scenario bands. To avoid reading PUE as a pure efficiency metric, the scenario design treats it as an equilibrium outcome shaped by cooling architecture and support energy substitution, consistent with how large facilities report operating improvements [35].

Electricity prices that used to value avoided cooling electricity are anchored to official U.S. monthly price series, using the average price of electricity to ultimate customers as a transparent benchmark [36]. Financial parameters, including the discount rate, inflation treatment, and levelization convention, follow standard assumptions used in techno-economic assessments and government-supported baselines, which keeps the results comparable with related infrastructure studies [37]. When needed, all monetary values are converted to a common base year, as reported in Table 2.

(b) The WtE block is defined by the recoverable thermal stream available for export after internal needs are met. At the screening level, it is parameterized by municipal solid waste lower heating value (LHV), throughput (t/day), and an exportable thermal fraction that captures boiler performance, internal heat use, and any operational commitment of heat to electricity generation. Typical heat-balance magnitudes and plant-level operating envelopes are anchored to established WtE engineering references and synthesis studies on practical ranges for energy recovery and exportable heat streams [38,39]. The analysis distinguishes two export forms that matter for coupling feasibility: (i) steam export and (ii) hot-water export suitable for district-energy delivery. The temperature band is listed explicitly in Table 2 because absorption performance depends on driving temperature. Emissions and sustainability accounting follow a reporting boundary aligned with established inventory guidance for waste-treatment and incineration categories, which allows consistent ESG translation without attempting a full life-cycle optimization [40]. The paper therefore does not infer a unique power-to-heat ratio from an explicitly specified turbine configuration such as back-pressure or extraction-condensing operation. Rather, these engineering differences are bracketed at screening level through the net exportable thermal fraction, interpreted as the heat stream available for external cooling service after internal uses, cycle configuration, and extraction priorities are accounted for.

(c) Thermal-to-cooling conversion is represented by the absorption COP and the associated driving temperature band. In the main screening results, the reference configuration is a single-effect LiBr absorption chiller, consistent with the hot-water-equivalent driving-heat band reported in Table 2 [41]. Double-effect systems are acknowledged in the literature and retained in Appendix A.3 for context, but they are not used as the baseline in the main figures because they imply a different, higher-grade heat regime and a different infrastructure configuration than the corridor-scale WtE coupling studied here.

(d) The coupling corridor is represented as a trunk delivery path of length $L$ (km). At the screening stage, the two main infrastructure constraints are thermal loss and pumping parasitics. Delivered driving heat is modeled as $\dot{Q}_{del}(L) = \dot{Q}_0 \exp(-\beta L)$, where $\beta(1/\text{km})$ is a reduced-form loss rate. To calibrate $\beta$, published pre-insulated pipe handbooks are used to extract linear heat-loss values (W/m) by temperature difference and insulation standard, then convert them into fractional loss per kilometer under a representative trunk thermal flow [32]. District-energy reviews are used to check whether the implied rates remain plausible under modern low-temperature network designs and corridor practices [31]. Pumping is parameterized either as a specific electric intensity, such as kWh per MWh of heat delivered, or as a fraction of delivered thermal power under the assumed flow regime. Recent thermo-hydraulic district heating studies provide order-of-magnitude ranges and show how hydraulic design and network temperature levels affect parasitic electricity demand [33]. Seasonal ground-temperature variation is not separately simulated as a transient soil-heat-transfer problem; instead, its effect is conservatively absorbed into the corridor loss-rate band together with insulation quality and effective temperature differential. Likewise, network pressure-drop effects are not modeled as a routed hydraulic solution. Their impact is absorbed into the pumping-intensity term and the scenario band for parasitic burden, which is intended to bracket plausible burdens under different routing, diameter, temperature-level, and hydraulic-station conditions.

(3) Scenario construction and sensitivity design

Three scenario packages are used to separate physical feasibility uncertainty from economic context and climate-boundary conditions. The conservative case is intentionally unfavorable to coupling, with higher thermal decay, higher pumping parasitics, lower absorption COP, and a lower electricity price. The aggressive case represents favorable but literature-consistent conditions, with lower decay, lower parasitics, higher COP, and a higher electricity price. The baseline case is set near literature midpoints. These packages are summarized in Table 3. Comparative statics are then conducted for the main drivers of corridor feasibility and grid relief: IT utilization $u$, cooling coverage fraction, coupling distance $L$, waste LHV and throughput, absorption COP (linked to temperature), and parasitic electricity intensity. This design yields a siting-ready viable frontier that can be read as a planning rule. Coupling is attractive where avoided baseline cooling electricity exceeds delivery losses and parasitic electricity, being less robust outside that range. The parameter and scenario tables are structured so that project teams can replace the screening assumptions with site data, such as routing, diameter, hydraulic stations, dispatch constraints, and permitting conditions, while retaining the same accounting structure [41].

**Table 2.** Core parameters and scenario values (used for Figures 3–6 and sensitivity bands).

| Parameter | Symbol | Unit | Consv | Baseline | Aggres | Definition/How Used | Anchoring Notes |
|---|---|---|---|---|---|---|---|
| MSW lower heating value (as received) [25] | $LHV_{\text{MSW}}$ | MJ/kg | 8 | 10 | 12 | Feedstock energy content driving recoverable heat | Reported MSW LHV commonly ~8–12 MJ/kg; average ~10 MJ/kg appears frequently in the MSW/WtE literature. |
| WtE throughput (representative plant) [25,31] | $\dot{m}_{\text{MSW}}$ | t/day | 500 | 1500 | 3000 | Scales available chemical energy and thermal export | Large facilities commonly exceed hundreds to thousands t/day; used only as scaling input for $Q_{\text{drive,0}}$(regulatory threshold context; plant scale examples widely reported). |
| Exportable thermal fraction (net) [25,31] | $\eta_{\text{th,export}}$ | - | 0.30 | 0.45 | 0.60 | Converts MSW chemical energy into exportable driving heat stream | Screening parameter capturing boiler efficiency, internal loads, and export share; reported here as scenario lever; users can replace with plant-specific heat balance. |
| Reference steam conditions (for context) [30,31] | - | bar, °C | 40 bar, 400 °C | 40 bar, 400 °C | 40 bar, 400 °C | Context for "modern WtE" steam availability; not a decision variable | Typical WtE steam conditions often cited around 40 bar, 400 °C; higher parameters exist with corrosion control. |
| Driving heat temperature band (hot-water equivalent) [31–33] | $T_{\text{drive}}$ | °C | 80 | 90 | 110 | Determines feasible absorption configuration and COP band | Single-effect LiBr commonly requires ~80–100 °C driving heat. |
| Absorption COP (single-effect band) [18,32] | $COP_{\text{abs}}$ | - | 0.65 | 0.75 | 0.85 | Converts delivered driving heat into delivered cooling | Published single-effect COP typically ~0.6–0.8; values ~0.7–0.8 commonly reported. |
| Climate sink proxy (cooling-water temperature) [34,35] | $T_{\text{cw}}$ | °C | 30 | 27 | 24 | Affects achievable COP and auxiliary power; implemented as scenario boundary | Represents hot/humid vs. moderate vs. cool/dry bands; can be tied to wet-bulb + approach if desired. |

| Pipe linear heat loss (modern insulated) [32] | $q_\ell$ | W/m | 40 | 25 | 10 | Intermediate anchor used to derive $\beta$ band | Pipe guides and examples report heat loss in W/m; 10–40 W/m is representative across ΔT and insulation classes. The values are intended to bracket plausible corridor losses under modern pre-insulated piping rather than to represent a project-specific buried-pipe thermal simulation. |
|---|---|---|---|---|---|---|---|
| Representative trunk thermal flow (for mapping) [18] | $Q_{\text{trunk}}$ | $MW_{th}$ | 2 | 5 | 10 | Used to translate $q_\ell$into $\beta$via $\beta \approx (q_\ell \cdot 1000)/Q_{\text{trunk}}$ | Makes explicit that fractional loss that depends on load level; can be replaced with plant/network design flow. |
| Thermal decay rate (implied) [32] | $\beta$ | 1/km | 0.020 | 0.005 | 0.001 | Used in $Q_{\text{drive,del}} = Q_{\text{drive,0}}\exp(-\beta L)$ | These values correspond to the $q_\ell$ and $Q_{\text{trunk}}$ anchors above; reported as the reduced-form corridor decay parameter. |
| Pumping electricity intensity [33] | $e_{\text{pump}}$ | $\text{kWh}_e/\text{MWh}_{th}$ | 10 | 6 | 2 | Converts delivered thermal flow into required pumping electricity | Recent district heating modeling reports values around several $\text{kWh}_e/\text{MWh}_{th}$ (e.g., ~6 in 4GDH cases). |
| Absorber-plant auxiliary fraction (including cooling-tower fan/pump equivalents) [25,26,29] | $\alpha_{\text{aux}}$ | % of $Q_{\text{cool}}$ (as electric equiv.) | 6% | 4% | 2% | Captures absorber auxiliaries, cooling-tower fan/pump equivalents, and control loads associated with heat rejection in the absorption cooling configuration. | Used to prevent "free cooling" artifacts; conservative case penalizes auxiliaries. |
| Baseline PUE (electric-only boundary) [34,35] | $PUE_{\text{base}}$ | - | 1.50 | 1.35 | 1.20 | Baseline facility electricity burden (IT + support) | Screening boundary condition; not claimed as ISO-comparable after service substitution. |
| Electricity price [36] | $p_e$ | \$/kWh | 0.06 | 0.12 | 0.18 | Used in LCOC and viable economics | Represents regional variation; conservative uses low price (less benefit from avoided electricity). |
| Discount rate (real) [37] | $r$ | % | 10% | 7% | 5% | Used in annualization of CAPEX in LCOC | Screening financial assumption; replaceable with project WACC. |
| Pipe network CAPEX (installed, supply+return, inclusive) | $c_{\text{pipe}}$ | \$/m | 1200 | 750 | 500 | Used to translate distance into CAPEX penalty | District heating distribution network construction cost is around \$750/m for typical network diameters and is reported in techno-economic modeling literature; we adopt scenario bands around this anchor. |

Notes: Notation "C/B/A" = Conservative/Baseline/Aggressive, where a parameter is derived (not directly assumed); we report the derivation mapping.

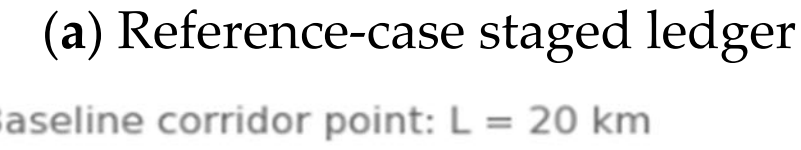
(**a**) Reference-case staged ledger

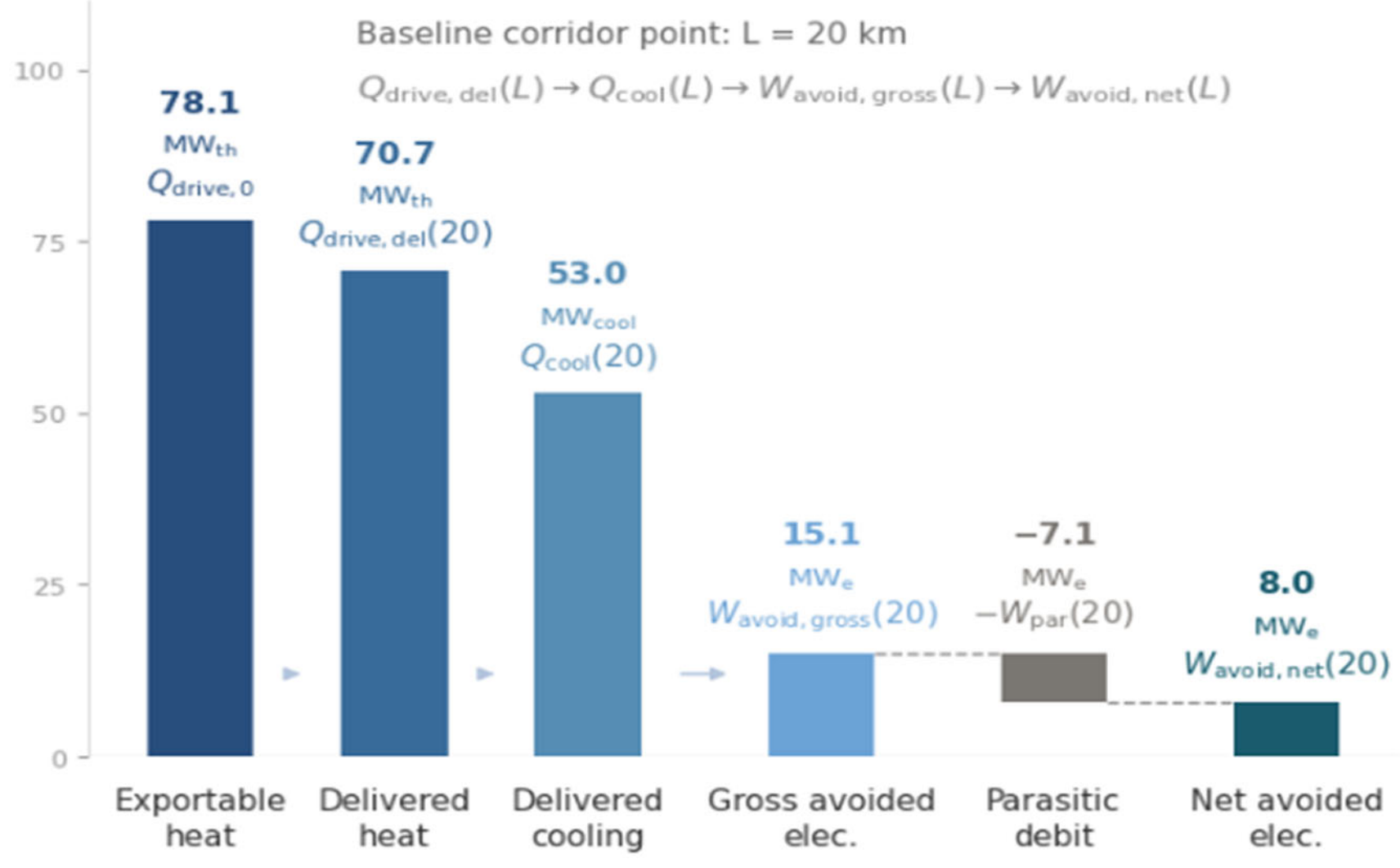

(**b**) Distance-point translation

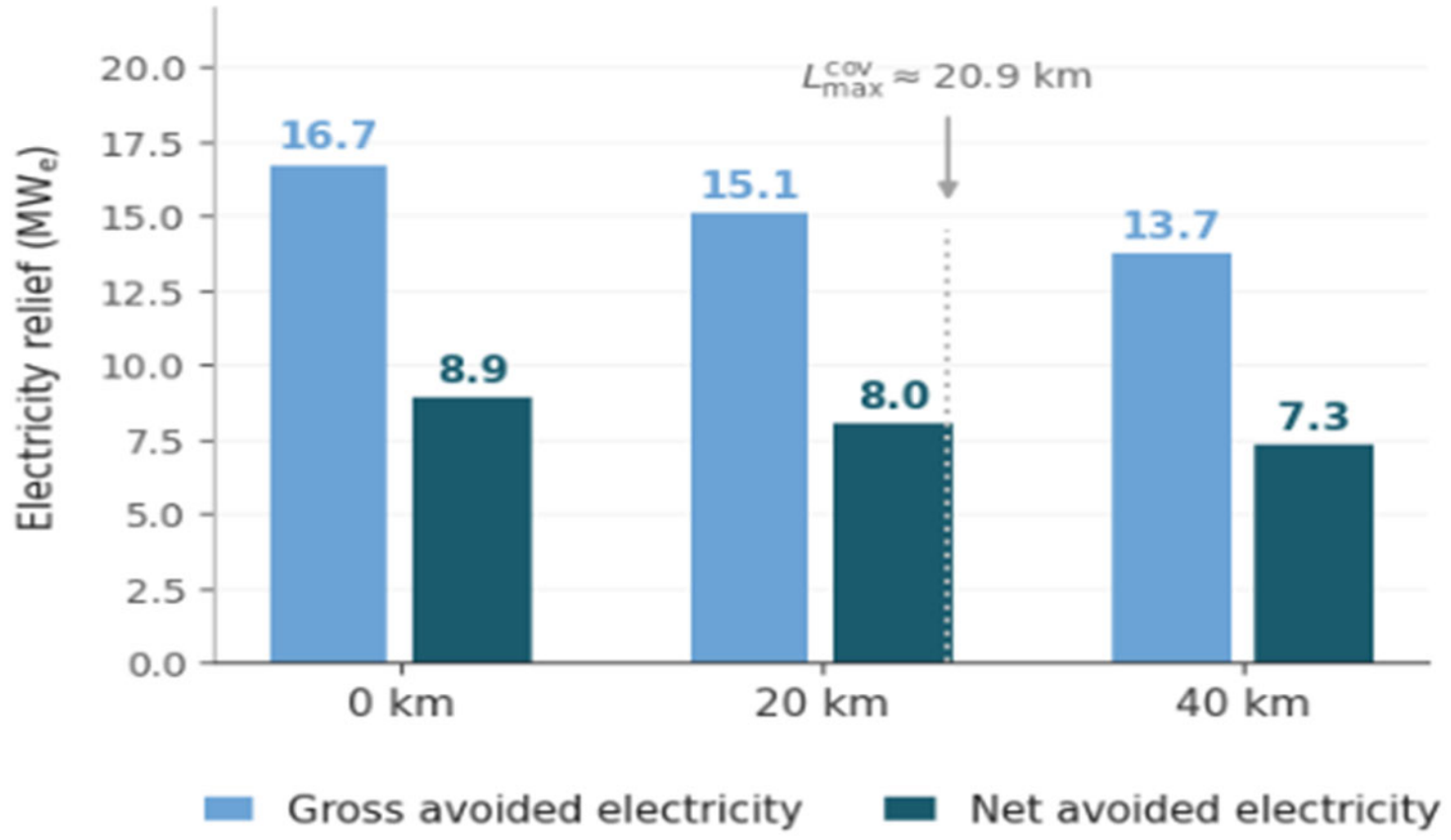

**Figure 3.** Reference-case energy-service ledger and corridor translation. Notes: $Q_{drive,0}$ denotes exportable driving heat available at the WtE plant boundary after internal plant-side uses and commitments are absorbed into the screening parameterization. $Q_{\mathrm{drive,del}}(L)$ is the delivered driving heat after corridor attenuation over distance $L$, where $L$ is the WtE–AIDC separation distance in km and $\beta$ is the reduced-form corridor decay parameter. $Q_{\mathrm{cool}}(L)$ is the delivered cooling service obtained from the absorption conversion of the delivered driving heat, with $COP_{\mathrm{abs}}$ denoting the absorption chiller coefficient of performance. $W_{\mathrm{avoid,gross}}(L)$ is the gross electric-equivalent displacement of the benchmark mechanical-chilling system, computed using the benchmark mechanical-cooling coefficient of performance $COP_m$. $W_{\mathrm{par}}(L)$ denotes parasitic electricity associated with corridor pumping and absorber-side auxiliaries. $W_{\mathrm{avoid,net}}(L)$ is the net avoided cooling electricity after parasitic debiting, i.e., $W_{\mathrm{avoid,net}}(L) = W_{\mathrm{avoid,gross}}(L) - W_{\mathrm{par}}(L)$. Units are reported as $\mathrm{MW}_{th}$ for thermal driving heat, $MW_{cool}$ for delivered cooling, and $\mathrm{MW}_e$ for electricity. The marker $L_{\max}$ denotes the coverage-defined threshold distance at which the coupled configuration can just fully cover the illustrative baseline cooling-electricity requirement under the reference parameterization. In panel (a), the light arrows indicate the staged translation from exportable heat to delivered heat, delivered cooling, gross avoided electricity, parasitic debit, and net avoided electricity; in panel (b), the dashed outline highlights the 20 km reference-point net avoided electricity corresponding to the baseline corridor case in panel (a).

**Table 3.** Scenario packages used for primary figures.

| Package | Engineering (Loss/COP/Pumping) | Climate Sink | Economic Context | Intended Interpretation |
|---|---|---|---|---|
| Conservative | High $\beta$, low $COP_{\text{abs}}$, high $e_{pump}$, higher $\alpha_{\text{aux}}$ | Hotter sink (higher $T_{\text{cw}}$) | Lower $p_e$, higher $r$, higher $c_{pipe}$ | Lower-bound feasibility; "corridor shrinks quickly" under unfavorable but plausible conditions |
| Baseline | Midpoint $\beta$, midpoint $COP_{\text{abs}}$, midpoint $e_{pump}$ | Moderate sink | Midpoint prices/discount/cost | Representative screening case |
| Aggressive | Low $\beta$, higher $COP_{\text{abs}}$, low $e_{pump}$, low $\alpha_{aux}$ | Cooler sink | Higher $p_e$, lower $r$, lower $c_{pipe}$ | Upper-bound feasibility envelope; still literature-consistent |

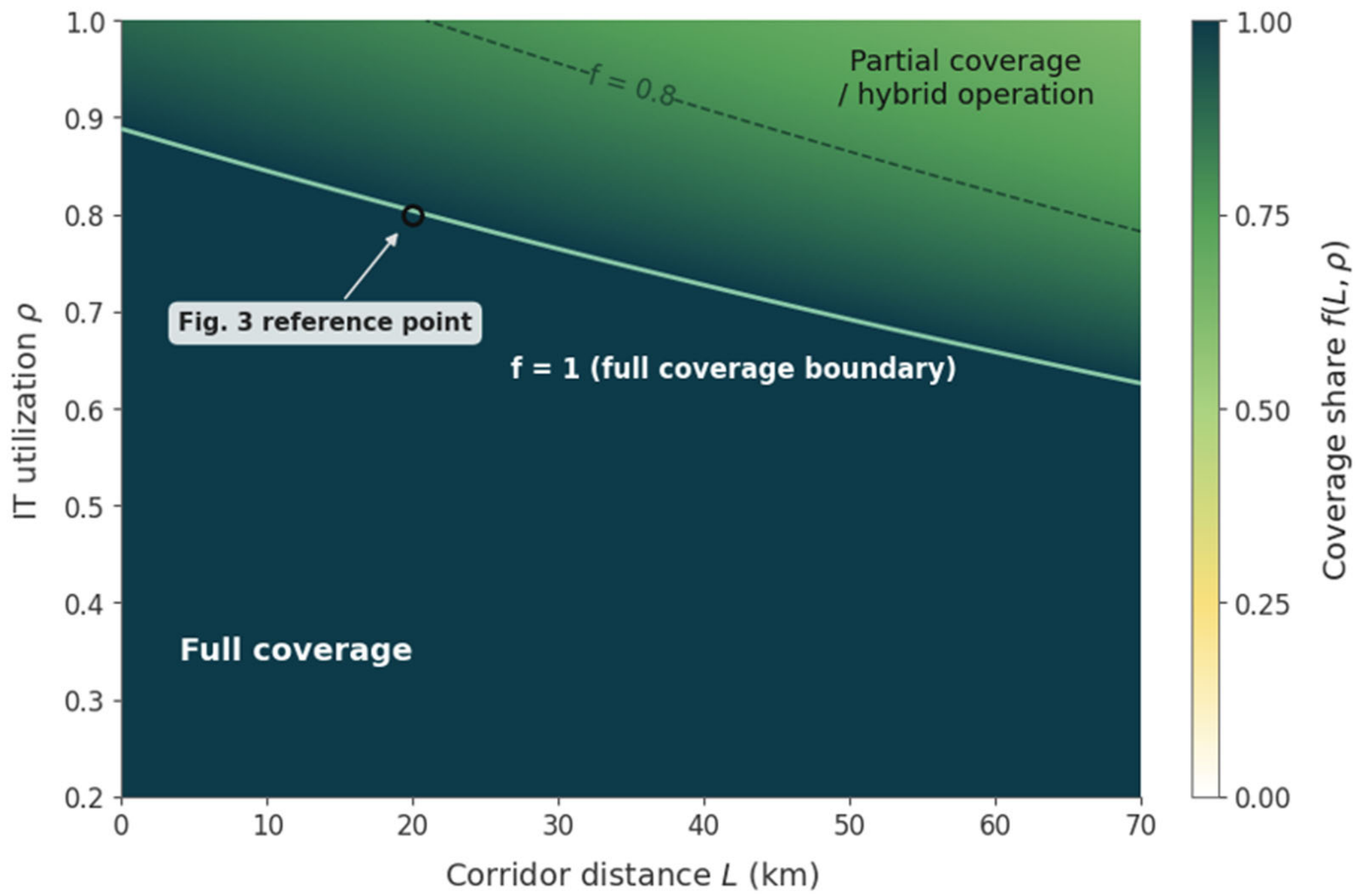

**Figure 4.** Coverage regimes under corridor separation and IT utilization. Notes: $L$ denotes corridor distance (km), $\rho$ denotes IT utilization, and $f(L,u)$ denotes the heat-driven coverage share, defined as $f(L,u) = \min\{1, Q_{\text{cool}}(L)/Q_{\text{req}}(u)\}$. Here, $Q_{\text{cool}}(L)$ is delivered cooling from the WtE-coupled absorption pathway after corridor attenuation, and $Q_{\text{req}}(\rho)$ is the cooling requirement associated with the operating IT load. The region below the $f = 1$ contour corresponds to full coverage, where residual mechanical cooling is zero under the stated bookkeeping boundary; the region above it corresponds to partial coverage or hybrid operation, where a remaining share of cooling demand must still be met mechanically. The figure uses the same baseline reference-case family as Figure 3.

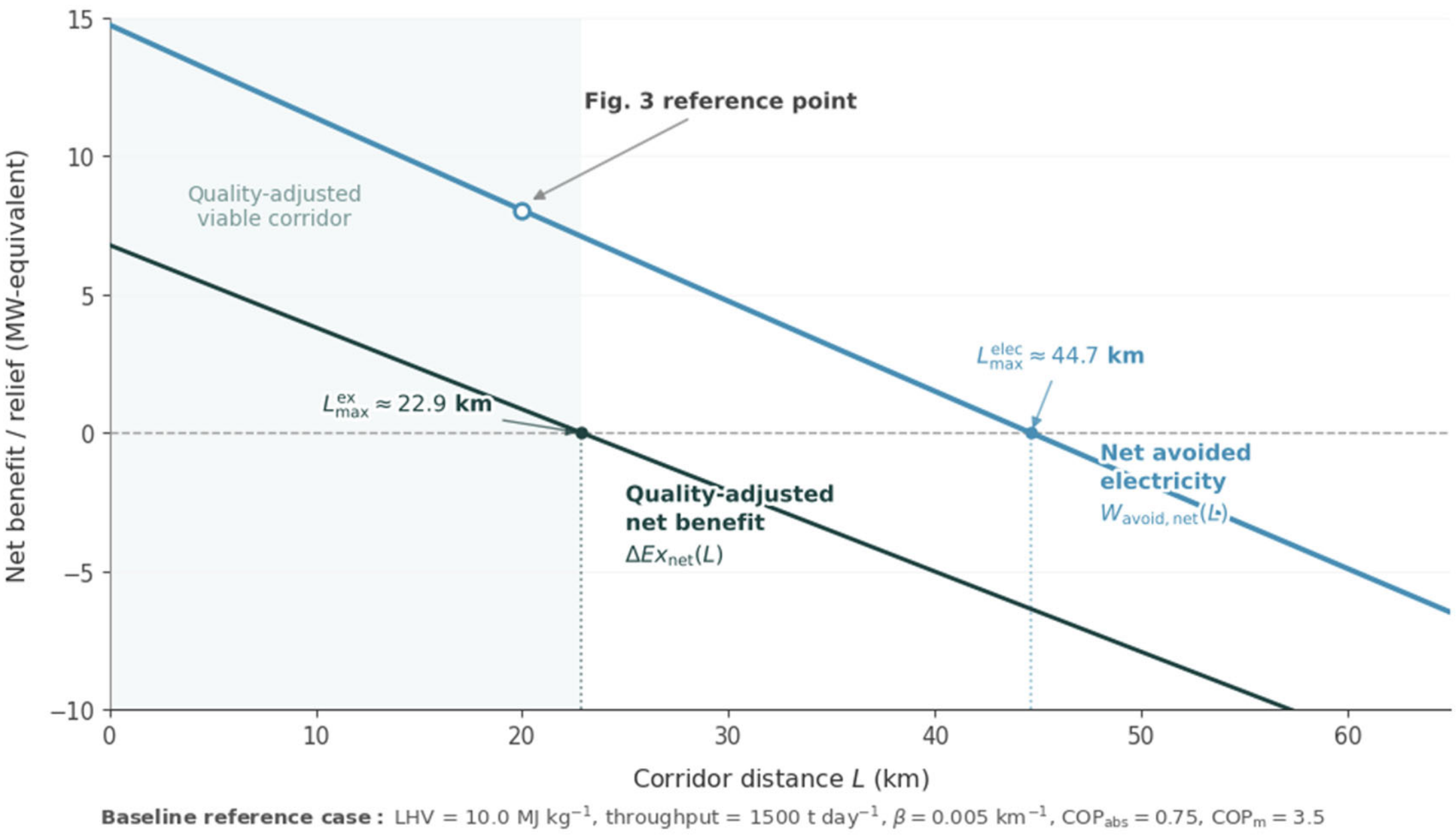

**Figure 5.** Cooling coverage and electricity displacement under feedstock variability: sensitivity to waste LHV and throughput. Notes: $L$ denotes corridor distance. $W_{\text{avoid,net}}(L)$ is net avoided cooling electricity after parasitic debiting. $\Delta Ex_{\text{net}}(L)$ is the quality-adjusted net benefit, which credits delivered cooling by its exergy factor $\phi_c$ and debits parasitic electricity within the same accounting boundary. $Q_{\text{cool}}(L)$ decreases with corridor attenuation, while $W_{\text{par}}(L)$ remains inside the corridor accounting boundary. Positive values indicate that the corresponding criterion remains favorable under the stated screening assumptions. The figure uses the same baseline reference-case family as Figures 3 and 4. The shaded region indicates the corridor interval that remains viable under the stricter quality-adjusted criterion, i.e., where $\Delta Ex_{net}(L) \geq 0$.

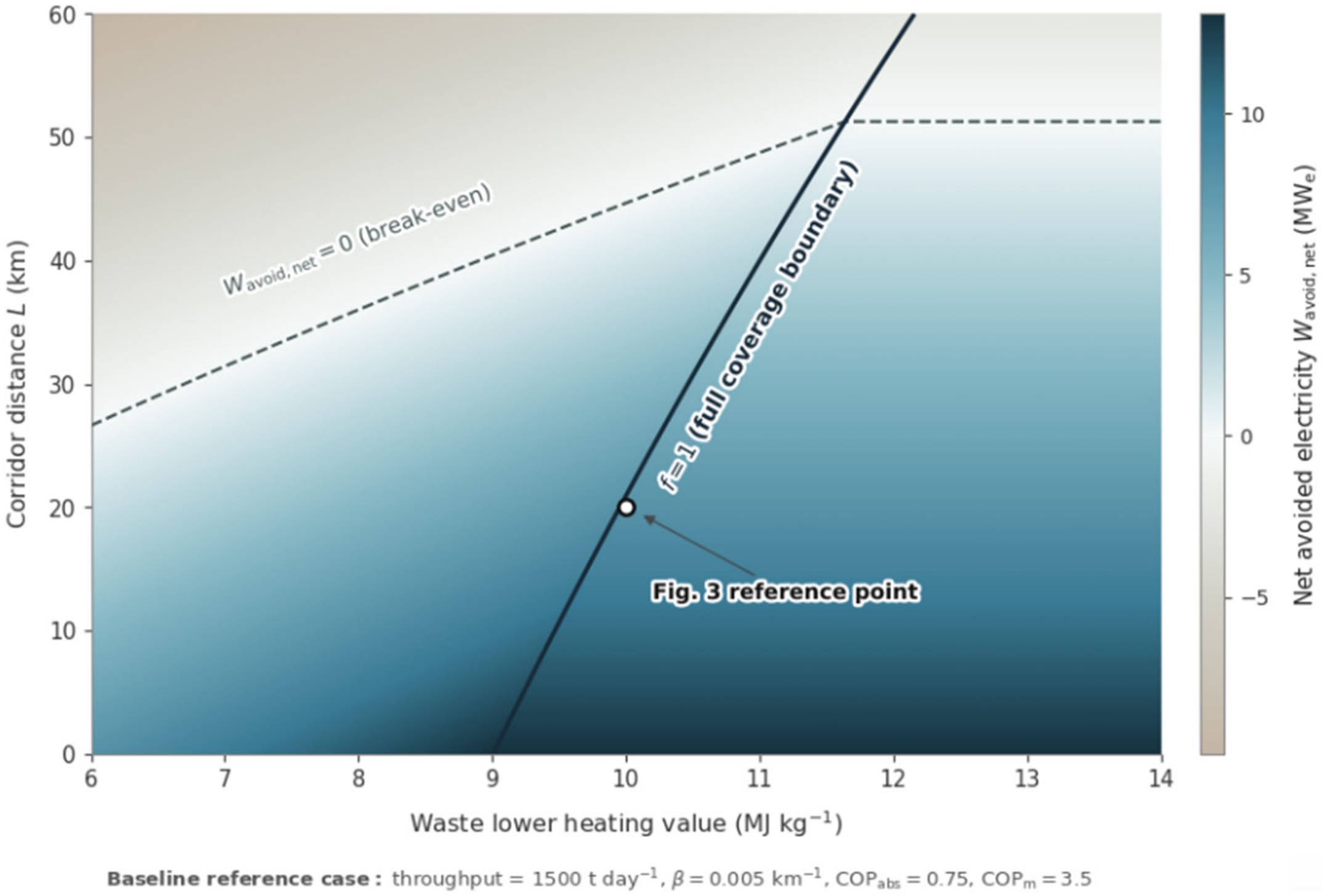

**Figure 6.** Feedstock quality, corridor distance, and net avoided electricity relief. Notes: The heatmap reports net avoided cooling electricity, $W_{\text{avoid,net}}$, as a joint function of waste lower heating value (LHV) and corridor distance $L$. The solid contour $f = 1$ marks the full-coverage boundary; below it, the baseline cooling-electricity requirement is fully displaced under the stated bookkeeping boundary. The dashed contour $W_{\text{avoid,net}} = 0$ marks the electricity–relief break-even condi-

tion. The region between the two contours represents partial coverage with still-positive net electricity relief. The marker denotes the Figure 3 reference point ($LHV - 10MJkg^{-1}, L - 20km$). All other parameters follow the baseline reference-case family.

## 3. Screening Results and Deployment Translation

This section develops the screening results from a reference-case ledger to corridor-level planning objects. It then identifies operating regimes, quality-adjusted feasibility thresholds, and scenario-specific corridor bands, before translating those results to deployment scale for a 1 GW IT campus.

### *3.1. Reference-Case Energy-Service Ledger*

Before turning to the comparative statics, we instantiate the screening framework with a single midpoint parameter set drawn from Appendix A.3. The purpose is not to claim site-specific feasibility, but to show how the model translates observable inputs—WtE exportable heat, corridor attenuation, absorption conversion, and parasitic electricity—into the paper's planning objects: net avoided cooling electricity and a coverage-defined corridor threshold. This reference configuration is therefore a quantified screening instantiation rather than a realized project case. Within this screening boundary, the formulation remains first-principle-based in the accounting sense: plant boundary exportable heat is translated into delivered cooling and then into gross and net electricity relief through explicit attenuation, thermal conversion, and parasitic debiting, rather than through a collapsed efficiency factor.

Consider a representative modern regulated WtE facility with municipal solid-waste lower heating value of 10 $\mathrm{MJ\ kg^{-1}}$ and a throughput of 1500 $\mathrm{t\ day^{-1}}$. Under the baseline net exportable thermal fraction, the plant provides approximately 78.1 MW$\mathrm{MW}_{th}$ of driving heat at the coupling boundary. Within the accounting boundary of this paper, that quantity is interpreted as exportable heat available for external cooling service after internal uses and plant-side commitments have been absorbed into the screening parameterization.

The reference-case translation follows a compact screening ledger:

$$Q_{\mathrm{drive,del}}(L) = Q_{\mathrm{drive,0}} \exp(-\beta L), Q_{\mathrm{cool}}(L) = COP_{\mathrm{abs}}\, Q_{\mathrm{drive,del}}(L) \quad (17)$$

$$W_{\mathrm{avoid,gross}}(L) = \frac{Q_{\mathrm{cool}}(L)}{COP_m}, W_{\mathrm{avoid,net}}(L) = W_{\mathrm{avoid,gross}}(L) - W_{\mathrm{par}}(L) \quad (18)$$

where the midpoint reference case uses $\beta = 0.005\mathrm{km}^{-1}$ (corridor decay), $COP_{\mathrm{abs}} = 0.75$ (single-effect absorption), and $COP_m = 3.5$ (absorption COP). In this way, corridor deliverability, absorption conversion, benchmark electric displacement, and parasitic debiting are linked through a single auditable energy-service chain rather than through separate descriptive steps. Here, $Q_{drive,0}$ denotes the plant boundary net exportable heat introduced in Figure 1 accounting topology.

Figure 3a makes this accounting chain explicit at the baseline corridor point $L = 20$ km. Starting from 78.1 $\mathrm{MW}_{th}$ of exportable driving heat at the plant boundary, corridor attenuation reduces the delivered driving heat to 70.7 $\mathrm{MW}_{th}$, which in turn yields 53.0 $\mathrm{MW}_{cool}$ of delivered cooling under the baseline absorption conversion. When this cooling service is valued against the benchmark mechanical-chilling system, the corresponding gross avoided electricity is 15.1 $\mathrm{MW}_e$. After deducting 7.1 $\mathrm{MW}_e$ of corridor pumping and absorber-side auxiliary electricity, the resulting net avoided cooling electricity is 8.0 $\mathrm{MW}_e$. The staged ledger is useful precisely as it prevents the reference case from collapsing into a single net number: it shows separately how much value is created by delivered cooling and absorbed by transport and auxiliary penalties within the stated boundary.

Figure 3b then translates the same reference configuration across three corridor distances. At $L = 0$ km, the implied net avoided cooling electricity is approximately 8.9 MW; at $L = 20$ km, it is 8.0 MW; and at $L =$ 40 km it remains 7.3 MW. Expressed in this form, the reference case does not present a single "best" point estimate. Instead, it provides a transparent distance-based ledger, showing that the benefit remains materially positive over short-to-moderate corridors, while also making visible the monotonic erosion caused by spatial separation. This is the sense in which the reference case is informative for planning: it converts a WtE thermal stream into a campus-relevant electricity–relief object without hiding the intermediate losses.

To express the same configuration in corridor-planning terms, we define an illustrative AIDC bookkeeping target with an IT power of 40 MW and baseline cooling-electricity demand of 8 MW. Under the baseline parameterization, full thermal coverage is maintained up to approximately $L_{\max} = 20.9$ km, as also indicated in Figure 3b. This threshold is the key planning output produced by the reference case: it converts the coupled energy-service ledger into a distance-based feasibility marker that can be compared directly across the later sensitivity results. The same is true for plant-cycle-specific heat extraction and explicit pressure-drop calculations in the thermal corridor, both of which are intentionally deferred to project-calibrated design studies while preserving a transparent screening-level accounting structure in the present paper. Instead, external validity is addressed here by anchoring the first-order drivers to literature-based ranges and by expressing the corridor frontier in terms of a small set of replaceable parameters.

*3.2. Coverage Regimes, Operating Envelope, and Electric Displacement*

Building on the reference-case ledger in Section 3.1, the next question is how the coupled configuration behaves across the joint operating space of corridor separation and IT utilization. In the present screening framework, MSW variability enters through feedstock LHV, throughput, and the net exportable thermal fraction rather than through a short-interval dispatch simulation. The resulting interpretation is hybrid rather than exclusive: WtE-driven absorption cooling displaces a share of baseline cooling electricity, but temperature-critical campus cooling service remains stabilizable through the residual mechanical-cooling path when delivered thermal service is reduced by lower feedstock quality, lower throughput, or corridor losses. For a screening paper, this regime map is more informative than a single efficiency curve because it shows where the heat-driven pathway can fully cover baseline cooling demand and where the system necessarily reverts to hybrid operation with residual mechanical chilling.

In the present framework, $Q_{\text{cool}}(L) = COP_{\text{abs}}\, Q_{\text{drive,0}} \exp(-\beta L)$, while the cooling requirement rises with operating IT load, $Q_{\text{req}}(u)$, where $u \in (0,1]$ denotes utilization. We therefore define the heat-driven coverage ratio as $f(L,u) = \min\left\{1, \frac{Q_{\text{cool}}(L)}{Q_{\text{req}}(u)}\right\}$. Figure 4 maps this object directly. The horizontal axis is corridor distance $L$, the vertical axis is IT utilization $\rho$, and the color scale reports the share of baseline cooling demand covered by the WtE-driven absorption pathway. The solid contour $f = 1$ separates the full-coverage regime from the partial-coverage regime, while the lighter dashed contour highlights a deeper partial-coverage zone. The marker denotes the Figure 3 reference point, linking the single-point ledger in Section 3.1 to the broader operating envelope.

Three results follow. First, full coverage is concentrated where corridor separation is short and campus utilization is moderate, because delivered cooling remains large relative to the required cooling load. Second, the full-coverage region contracts monotonically as either distance increases or utilization rises: distance reduces delivered thermal service, while higher utilization raises cooling demand in absolute terms. Third, the regime boundary is itself a planning object, because it identifies which siting-load combinations remain compatible with full thermal coverage under the stated screening assumptions. The Figure 3 reference point lies close to this transition, which clarifies why a 20 km corridor is a useful baseline for the reference-case translation.

This operating-envelope view also clarifies the electric-displacement logic more directly than the earlier PUE-style representation. When $f(L,u) = 1$, residual mechanical cooling falls to zero, so further increases in thermal availability no longer reduce electric cooling demand. The electric benefit therefore plateaus once the system enters full coverage. When $f(L,\rho) < 1$, the coupled configuration remains hybrid: a share $1 - f(L,u)$ of the cooling requirement must still be met by mechanical chilling, and the electric benefit is only partial. The same regime logic carries directly to electric PUE. Once $f(L,u) = 1$, residual cooling electricity vanishes and further PUE improvement is structurally limited. In this sense, Figure 4 is not an efficiency scorecard but an operating-regime map. It shows when coupling remains full-coverage, when it becomes hybrid and partial, and why later gains are structurally bounded once the system has crossed into the full-coverage region.

These ledger outputs provide the physical inputs for the decision metrics defined in Section 2.5, including the cost-normalized comparison through LCOC.

3.3. Quality-Adjusted Benefit as a Corridor Object

Section 3.2 mapped the operating envelope of the coupled configuration by identifying where delivered thermal cooling fully covers the baseline cooling requirement and where the system remains hybrid. That regime map is necessary but not sufficient for thermoeconomic interpretation, because electricity relief alone does not fully characterize the value of converting low-grade heat into cooling service. A second criterion is therefore needed, one that credits delivered cooling by thermodynamic quality while explicitly debiting the high-grade electricity consumed by transport and auxiliaries. Within the present boundary, the quality-adjusted benefit is evaluated through $\Delta Ex_{\mathrm{net}}(L) = \phi_c\, Q_{\mathrm{cool}}(L) - W_{\mathrm{par}}(L)$, where $\phi_c$ is the cooling exergy factor, $Q_{\mathrm{cool}}(L)$ is delivered cooling, and $W_{\mathrm{par}}(L)$ is parasitic electricity. This formulation is conservative by construction: cooling is credited only by its exergy relative to ambient, while parasitic electricity is treated as an explicit debit within the same accounting boundary. In this sense, exergy is not introduced here as a separate thermodynamic layer for its own sake. It serves as the quality-adjusted bridge that allows delivered cooling and high-grade electricity use to be compared on a common basis.

The role of exergy here is deliberately limited but decision-relevant. Rather than resolving component-by-component irreversibilities within the WtE plant or absorption cycle, the present formulation uses exergy to quality-adjust delivered cooling at the corridor level, so that the thermal-service benefit and auxiliary electricity penalty remain comparable within a common screening boundary. This choice is conservative by construction: cooling is credited only by its exergy-equivalent value, while parasitic electricity is fully debited in the same bookkeeping frame.

Figure 5 reports the resulting corridor objects directly. The blue curve, $W_{\mathrm{avoid,net}}(L)$, tracks net avoided cooling electricity after parasitic debiting and therefore represents the electricity–relief criterion already introduced in Section 3.1. The darker curve, $\Delta Ex_{\mathrm{net}}(L)$, applies the stricter quality-adjusted criterion. Reading the two together is more informative than reporting a static exergy-efficiency number, because the figure shows not only that grade matching can be beneficial, but also how that benefit is compressed as corridor length increases and delivered cooling decays. The Figure 3 reference point at $L = 20$ km lies inside the positive region for both criteria, but it is visibly much closer to the quality-adjusted threshold than to the electricity–relief threshold. That geometry is informative in itself: under the baseline corridor assumptions, the reference case remains favorable, yet already sits near the stricter thermoeconomic limit.

Three results are central. First, both criteria remain positive over short corridors, confirming that the coupled pathway can remain attractive when thermal deliverability is high and parasitic burdens are modest. Second, both decline with distance, because attenuation reduces $Q_{\mathrm{cool}}(L)$ while parasitic requirements remain inside the accounting boundary. Third, the quality-adjusted criterion is more demanding. In Figure 5, $\Delta Ex_{\mathrm{net}}(L)$ crosses zero at approximately $L_{\max} \approx 22.9$ km, whereas the electricity–relief curve remains positive until roughly $L_{\max} \approx 44.7$ km. The gap between these two thresholds is analytically important. It shows that a corridor may still appear favorable under a pure electricity criterion while already looking marginal under the quality-adjusted criterion, because the latter credits only the exergy-equivalent share of delivered cooling rather than its full thermal magnitude.

Figure 5 should therefore be read as a corridor decision figure rather than as a summary efficiency comparison. Its role is to show how grade-matched cooling creates a measurable benefit at short distance, how that benefit is progressively compressed by corridor losses and parasitic work, and why quality-adjusted screening is more conservative than electricity relief alone. The shaded interval $0 \le L \le L_{\max}$ makes this interpretation explicit: it marks the corridor range that remains viable under the stricter quality-adjusted criterion, not merely under the electricity–relief criterion. This reading also prepares the transition to Section 3.5, where the associated threshold distances are stated explicitly as siting constraints rather than left implicit in the curve shapes.

*3.4. Feedstock Quality, Corridor Distance, and Net Electricity Relief*

Section 3.3 showed that corridor distance progressively compresses both electricity relief and quality-adjusted benefit. The next question is how supply-side variation modifies that corridor logic. In the present screening framework, feedstock uncertainty enters mainly through municipal solid-waste energy content, through-put, and the net exportable thermal fraction. Among these, lower heating value (LHV) is the most compact proxy for supply-side thermal quality at the reference-plant scale and is therefore used here as the main feedstock sensitivity variable. Figure 6 maps the net avoided cooling electricity, $W_{\text{avoid,net}}$, over the joint space of waste LHV and corridor distance. All other parameters follow the baseline reference-case family used in Sections 3.1–3.3, so the figure isolates how feedstock quality interacts with spatial deliverability. Read in this way, the heatmap is a planning surface rather than a generic parameter sweep.

Three results follow. First, net avoided electricity rises monotonically with feedstock quality, because higher LHV increases the thermal energy available at the plant boundary and enlarges the electric-equivalent displacement of baseline mechanical chilling. Second, net avoided electricity declines monotonically with corridor distance, because attenuation reduces delivered thermal service even when feedstock quality improves. Third, the interaction is nonlinear. At shorter corridors, moderate increases in LHV can materially raise relief and move the system rapidly toward full coverage. At longer corridors, the same increase yields a smaller gain because a larger share of the added thermal input is absorbed by distance-related decay.

The overlaid contours sharpen that interpretation. The solid $f = 1$ contour marks the onset of full coverage and shifts upward with distance, showing that longer corridors require better feedstock quality to maintain the same regime. The dashed $W_{\text{avoid,net}} = 0$ contour lies above it over the plotted range, so the system can leave full coverage before it loses positive net electricity relief. The region between the two contours therefore corresponds to hybrid operation with still-positive electricity benefit. Crossing the full-coverage boundary does not mean that the coupling ceases to be beneficial; it means that residual mechanical chilling re-enters the cooling mix. Only beyond the dashed contour does the electricity–relief criterion fail. The Figure 3 reference point $(LHV - 10MJkg^{-1}, L - 20km)$ lies essentially on the full-coverage boundary. This placement is informative. The baseline case is neither an extreme best case nor a marginally infeasible corner solution; it sits at the transition where additional feedstock quality still matters, but where marginal gains have already begin to narrow. Figure 6 should therefore be read as a planning surface. Better feedstock quality can widen the range of attractive WtE–AIDC pairings, but it does not remove the need for proximity.

*3.5. Scenario-Specific Threshold Bands Regarding Spatial Contraint*

Sections 3.1–3.4 developed the corridor logic progressively: the reference-case ledger translated WtE heat into electricity relief, the operating-envelope map separated full and partial coverage, the quality-adjusted criterion introduced a stricter corridor test, and the feedstock surface showed how supply-side conditions shift net relief. For planning, however, these continuous results must ultimately be reduced to threshold ranges. Figure 7 provides that synthesis by reporting scenario-specific corridor bands for the conservative, baseline, and aggressive packages.

Each scenario is summarized by two distances. $L_{\max}$ is the largest corridor length at which the heat-driven pathway can still fully cover the baseline cooling-electricity requirement under the stated bookkeeping boundary. $L_{\max}$ is the largest corridor length at which the quality-adjusted criterion remains non-negative. Together, these thresholds partition the corridor into three planning zones: full coverage $(L \leq L_{\max})$, partial but viable operation $(L_{\max} < L \leq L_{\max})$, and non-viable extension $(L > L_{\max})$.

Three results are central. First, the corridor narrows materially under conservative assumptions and widens under aggressive ones. Second, $L_{\max}$ and $L_{\max}$ move together but are not identical. The first is the operational threshold, marking the point at which residual mechanical chilling re-enters the cooling mix; the second is the stricter planning threshold, marking the point beyond which the coupled configuration no longer remains favorable under the quality-adjusted criterion. Third, the interval between them is decision-relevant in its own right. It identifies a corridor zone where the system is no longer fully heat-covered but still retains positive net benefit under the stricter screening rule.

Figure 7 also clarifies the status of the 20 km reference distance used in Figure 3. Under conservative assumptions, the same corridor already lies beyond overall viability. Under baseline assumptions, it sits essentially at the transition into partial operation. Under aggressive assumptions, it remains comfortably inside the full-coverage zone. This comparison is useful because it turns the reference case from a single numerical illustration into a benchmark that can be interpreted against a wider feasibility envelope.

Figure 7 should therefore be read as a planning synthesis rather than as another distance plot. For siting, the key question is not the exact shape of a single corridor curve, but whether a candidate pairing lies in the full-coverage zone, the partial-but-viable zone, or beyond the admissible corridor under plausible assumptions. Read in this way, the threshold bands provide the most compact statement of the paper's planning contribution: they convert the earlier continuous results into a directly interpretable corridor-feasibility map for WtE–AIDC pairing under uncertainty.

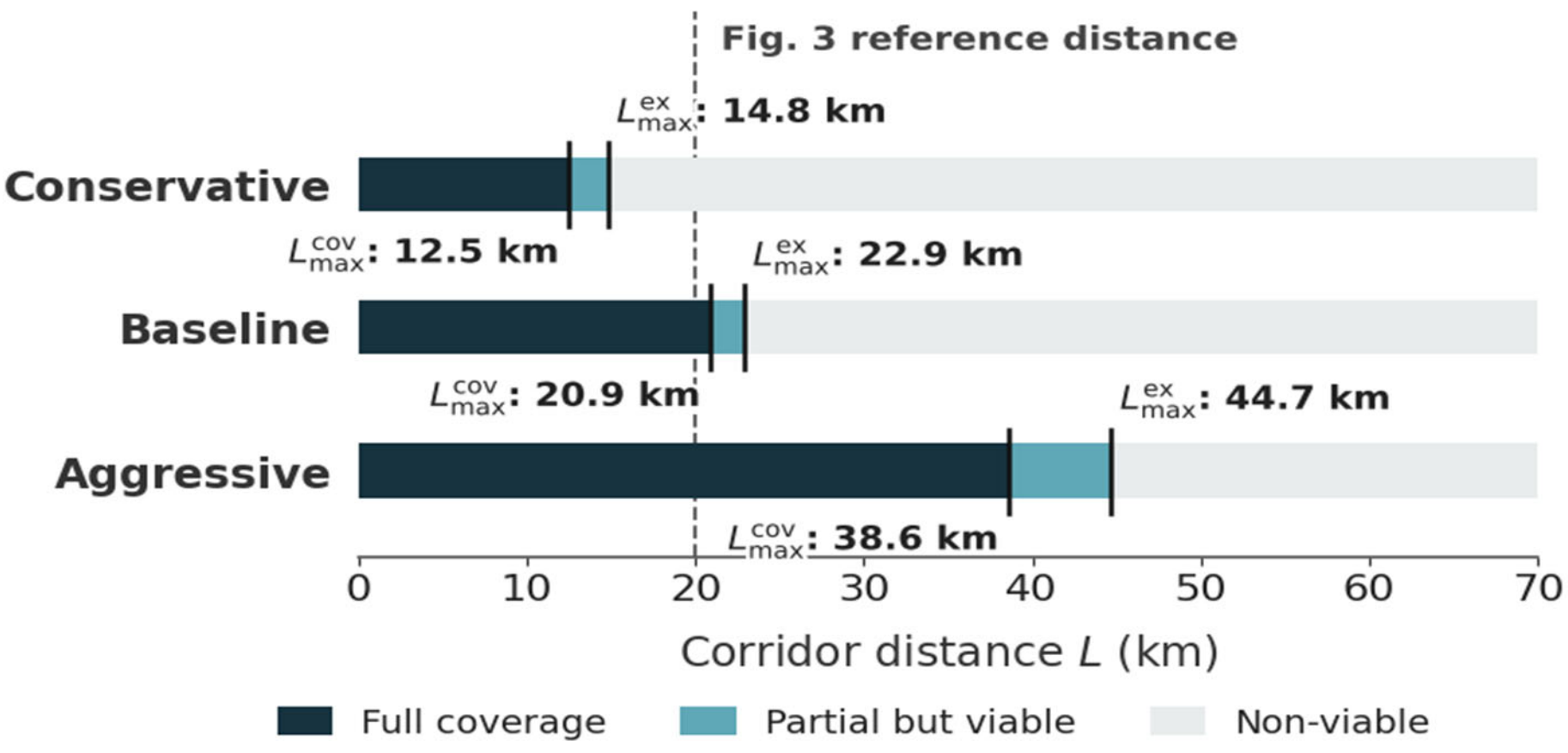

**Figure 7.** Scenario-specific corridor feasibility bands. Notes: Each row reports the scenario-specific full-coverage threshold $L_{\max}$ and quality-adjusted threshold $L_{\max}$. Together, these thresholds partition the corridor into three planning zones: full coverage ($L \le L_{\max}$), partial but viable ($L_{\max} < L \le L_{\max}$), and non-viable ($L > L_{\max}$). The dashed vertical marker denotes the Figure 3 reference distance ($L = 20$ km). Thresholds are derived from the conservative, baseline, and aggressive parameter packages.

Detailed plant-cycle modeling, component-resolved exergy destruction, heat-exchanger sizing, and pressure-drop calculations are intentionally deferred to project-calibrated engineering studies; the purpose of the present manuscript is to identify corridor-feasible planning regimes before such design-stage commitments are undertaken. The quality-adjusted and threshold-based corridor objects derived here are later used to interpret when LCOC-like economic comparisons remain meaningful under partial versus full coverage. In this sense, the scenario-specific corridor bands are the manuscript's primary scalability result: they do not assert universal deployability, but identify the distance ranges within which the coupled configuration remains operationally admissible and thermodynamically viable under stated parameter packages.

### *3.6. Deployment-Scale Translation (Per-GW Campus)*

This section maps the screening results into a deployment-scale planning lens for a 1 GW IT campus. The translation is designed to preserve the boundary logic of Section 3 while producing two quantities that are directly interpretable in infrastructure siting decisions: (i) campus-level grid-electricity relief under WtE-driven absorption cooling, and (ii) the implied scale of WtE infrastructure required to deliver a targeted displacement of cooling electricity across a finite corridor (the deployment translation is framed around AI data centers because the relevant constraint is not generic facility load, but large, sustained compute deployment under tight cooling and interconnection conditions). The full variable definitions, unit conversions, and the step-by-step computation protocol are documented in Appendix A.4.

#### 3.6.1. Objective and Planning Outputs

The deployment translation is not a GIS-based siting study and should not be read as evidence that the implied representative plant counts are physically available in the immediate vicinity of a fiber- and power-oriented AI campus. Rather, it is a scale-translation device that converts a representative WtE unit, a corridor distance, and a target displacement share into the amount of WtE-equivalent infrastructure required to offset a specified portion of baseline cooling electricity. Spatial availability of MSW feedstock, transfer logistics, permitting constraints, and compatibility with DC siting around fiber, power, and land nodes are outside the scope of this screening paper.

The deployment translation serves two purposes. First, it provides an auditable electricity accounting at the campus boundary that links the coupling channel to grid demand, consistent with standard facility reporting conventions. Second, it expresses feasibility as a corridor-planning problem by converting thermal deliverability into an equivalent representative WtE plant count as a function of coupling distance and desired coverage. Figure 8 reports the electricity ledger at a reference siting point, while Figure 9 reports the plant count surface over distance for multiple coverage targets. Both figures use the scenario packages in Tables 2 and 3 and adopt a 1 GW campus scale with utilization $u = 0.70$ and mechanical-cooling $COP_m = 4.0$.

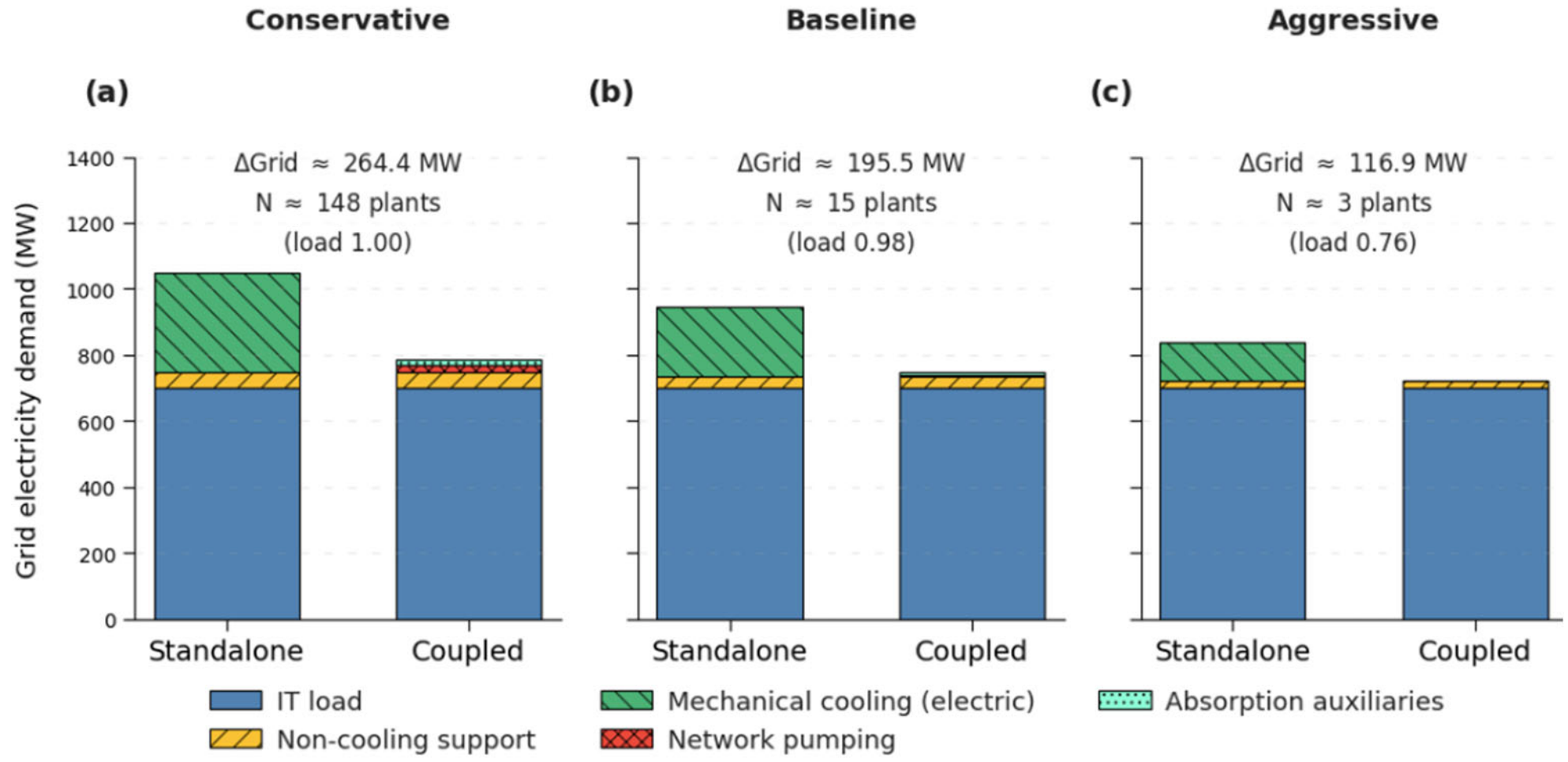

**Figure 8.** Per-GW AI campus grid-electricity accounting under WtE-driven absorption cooling (reference point). Notes: (a) Conservative package. Campus grid-electricity breakdown for the standalone and coupled configurations under the Conservative parameter package. (b) Baseline package. Campus grid-electricity breakdown for the standalone and coupled configurations under the Baseline parameter package. (c) Aggressive package. Campus grid-electricity breakdown for the standalone and coupled configurations under the Aggressive parameter package.
All panels use the same reference point, $u = 0.70$, coupling distance $L = 5.0$km, and mechanical cooling $COP_m = 4.0$. Stacked bars decompose grid demand into IT load, non-cooling support, and mechanical cooling electricity for the standalone benchmark. In the coupled case, mechanical cooling electricity is displaced by thermally driven absorption cooling, while network pumping and absorption auxiliaries are added as electricity penalties at the same boundary. $\Delta Grid$denotes the net reduction in grid electricity demand (MW) relative to the standalone benchmark after parasitics are included. $N$denotes the installed WtE plant count required to meet the full-coverage objective at the reference point, and “load” de-notes the implied average utilization of installed WtE-cooling capacity after integer siting.

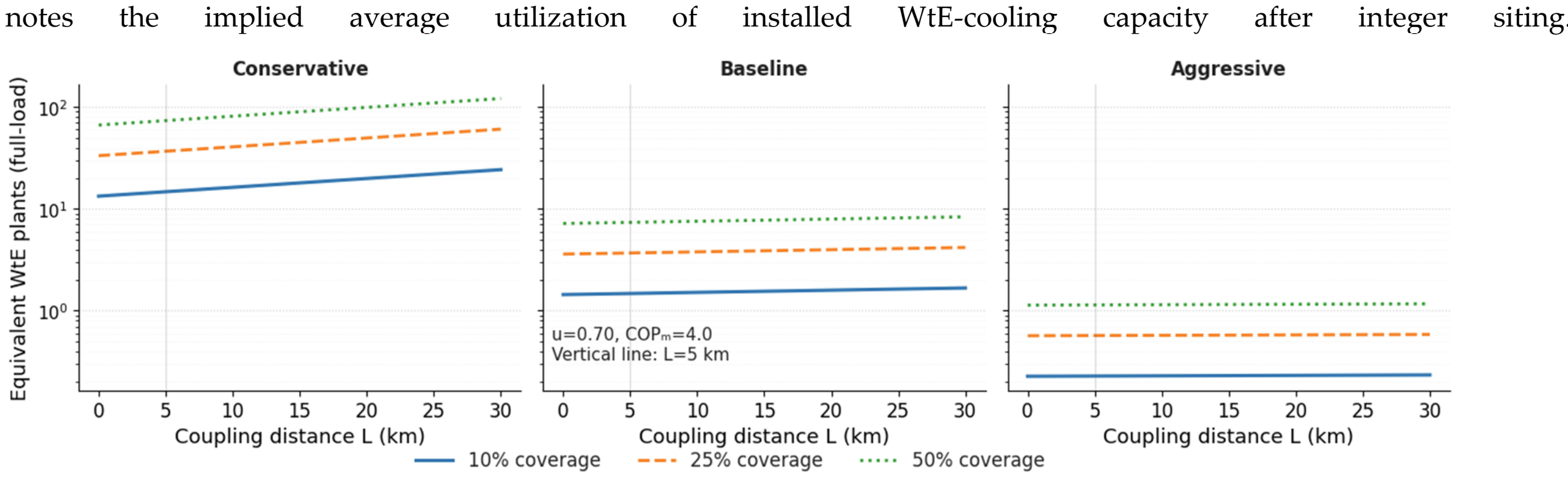

**Figure 9.** Equivalent WtE plants required to achieve target cooling-electricity displacement vs. coupling distance (per-GW campus). Notes: Curves report the full-load-equivalent (continuous) number of representative WtE plants required to displace $\phi = \{10\%, 25\%, 50\%\}$ of baseline mechanical-cooling electricity for a 1 GW IT campus under each scenario package (Tables 2 and 3). The campus scale uses $u = 0.70$ and $COP_m = 4.0$. Corridor deliverability is represented by exponential attenuation with decay rate $\beta$ (Table 2), so the required plant count rises with distance approximately as $e^{+\beta L}$. The vertical line marks $L = 5$ km, which is the reference point used for the deployment-scale accounting in Figure 8. Counts are reported as continuous equivalents to avoid rounding artifacts; discrete siting would round up to an integer and imply part-load operation when the continuous requirement is below the next integer.

### 3.6.2. Minimal Translation Definitions

Let $\bar{P}_{\mathrm{IT}}$ denote the campus IT nameplate and $u$ denote the utilization factor. Operating IT load is $P_{\mathrm{IT}} = u\,\bar{P}_{\mathrm{IT}}$. A baseline mechanical-cooling electricity quantity $W_{\mathrm{cool,base}}$ is defined for the standalone benchmark to serve as the displacement target. For a representative WtE plant, exportable driving heat $Q_{\mathrm{drive},0}$ is obtained from MSW throughput, heating value, and a net exportable fraction. Corridor deliverability is captured by exponential attenuation $e^{-\beta L}$, yielding $Q_{\mathrm{drive}}(L)$, which is converted into delivered cooling service via absorption $COP_{\mathrm{abs}}$. Net avoided grid electricity per plant, $W_{\mathrm{avoid,net}}(L)$, is defined as the gross avoided mechanical-cooling electricity net of pumping and auxiliary penalties. These objects determine both the campus electricity ledger (Figure 8) and the plant count required to achieve a target coverage share $\phi$ (Figure 9), via $N_{\mathrm{req}}(L;\phi)$ (Appendix A.4).

### 3.6.3. Translation Protocol (Summary)

The protocol proceeds in a single pass from Tables 2 and 3 to deployment figures. A scenario package is selected (Table 3), which fixes the parameter tuple in Table 2. Campus operating load $P_{\mathrm{IT}}$ is set by $(\bar{P}_{\mathrm{IT}}, u)$, and the standalone benchmark establishes $W_{\mathrm{cool,base}}$. For a given distance $L$, per-plant deliverable cooling and net avoided electricity are computed using the corridor decay and parasitic terms. Figure 8 evaluates this ledger at a reference point $L = 5$ km and reports the implied net grid relief. Figure 9 evaluates $N_{\mathrm{req}}(L;\phi)$ over $L \in [0,30]$ km for multiple $\phi$.

### 3.6.4. Scaled Results: Per-GW Electricity Accounting and Corridor Sizing

Figure 8 reports a deployment-scale grid-electricity ledger for a 1 GW IT campus at the reference point ($u = 0.70$, $L = 5$ km, $COP_m = 4.0$). Under this scaling, the campus IT load is $P_{\mathrm{IT}} = 700$ MW. The standalone bars decompose grid demand into IT load, non-cooling support, and mechanical-cooling electricity. The coupled bars show how much of the mechanical-cooling block can be displaced by thermally driven absorption cooling and the net of two electricity penalties at the same boundary: network pumping and absorption auxiliaries.

Baseline grid demand and its composition differ materially across packages. In the Conservative package, Standalone grid demand is approximately 1.05 GW, implying a non-IT overhead of roughly 350 MW above the 700 MW IT load. Within that overhead, the mechanical-cooling component is visually dominant, on the order of 250–270 MW, while non-cooling support is comparatively small, on the order of 60–70 MW. The Baseline package starts from a lower Standalone grid demand of roughly 0.95 GW, implying an overhead near 250 MW. Mechanical cooling remains the largest discretionary block, approximately 200–230 MW, with non-cooling support again on the order of 50–70 MW. The Aggressive package begins closest to the IT load, with the Standalone grid demand around 0.85 GW, implying an overhead near 150 MW. Mechanical cooling is materially smaller, about 120–140 MW, with non-cooling support around 30–50 MW.

Net grid relief is large in absolute terms and is largest where the baseline cooling-electricity block is largest. At the reference point, the coupled configuration reduces grid demand by $\Delta Grid \approx 264.4$ MW (Conservative), $\approx 195.5$ MW (Baseline), and $\approx 116.9$ MW (Aggressive). These values should be read as net effects after parasitics, not as a gross cooling-electricity displacement. Relative to the 700 MW IT load, the net relief corresponds to roughly 38% (Conservative), 28% (Baseline), and 17% (Aggressive) of IT power. Relative to total Standalone grid demand, the relief is on the order of one quarter to one third in the Conservative and Baseline cases, and still in the low-teens to mid-teens percent range in the Aggressive case. The non-monotonic ordering does not reflect weaker coupling physics in the Aggressive package. It reflects a smaller mechanical-cooling electricity burden available to displace in the Standalone benchmark.

The infrastructure burden required to obtain these relief levels is sharply heterogeneous. Figure 8 annotates the implied WtE plant counts required at the same reference point under the full-coverage sizing convention used to construct the Coupled ledger. The Conservative package requires approximately $N \approx 148$ plants, the Baseline package $N \approx 15$ plants, and the Aggressive package $N \approx 3$ plants. These orders of magnitude follow from the combined effect of (i) plant scale (throughput), (ii) exportable thermal fraction, (iii) absorption $COP_{\text{abs}}$, and (iv) corridor decay $\beta$, all drawn from Tables 2 and 3 and mapped through Appendix A.4. The implied "load" indicators, approximately 1.00 (Conservative), 0.98 (Baseline), and 0.76 (Aggressive), illustrate the discrete siting implication. When the required continuous equivalent is close to an integer, the fleet can operate near full load. When it is below the next integer, rounding creates part-load operation of the installed WtE-cooling capacity at the reference point, even though the campus-level displacement target is met.

Figure 9 generalizes the sizing question by plotting the full-load-equivalent number of representative WtE plants required to displace $\phi = \{10\%, 25\%, 50\%\}$ of the baseline mechanical-cooling electricity as a function of coupling distance $L \in [0,30]$ km. The vertical line marks $L = 5$ km, aligning the distance used in Figure 8 with the corridor sweep.

At the reference distance $L = 5$ km, the plant requirement spans two orders of magnitude across packages. In the Conservative package, required counts are approximately 15 plants (10% target), 40 plants (25% target), and 80 plants (50% target). In the Baseline package, the corresponding requirements are about 1.5 plants (10%), four plants (25%), and eight plants (50%). In the Aggressive package, they are approximately 0.25 plants (10%), 0.6 plants (25%), and 1.1 plants (50%). Even at the same distance and the same target definition, the implied infrastructure footprint ranges from "small fleet" to "multi-dozen" to "multi-site program," depending on the scenario package. Reporting full-load-equivalent counts makes this contrast transparent without forcing a discrete siting convention.

Distance sensitivity is meaningful but package-dependent, and it operates through corridor decay rather than through the cooling technology itself. Over 0–30 km, the Conservative curves rise notably: for example, the 50% target increases from roughly 70 plants near $L \approx 0$ to roughly 120 plants by $L \approx 30$ km. Baseline rises more modestly, with the 50% curve moving from roughly eight plants to roughly nine plants over the same range. Aggressive is nearly flat at all three target levels, with the 50% curve staying close to one plant.

Two additional quantitative readings follow directly from Figure 9. First, within each scenario package, raising the displacement target from 10% to 25% and then to 50% scale plant requirements almost proportionally. That proportionality indicates that, at these target levels, the system remains in a partial-coverage regime rather than being constrained by a hard cap in delivered cooling. Second, because the curves are plotted on a log scale, vertical separation between packages represents multiplicative differences in feasibility. At $L = 5$ km and $\phi = 25\%$, the Conservative package requires on the order of 40 plants, the Baseline package on the order of four plants, and the Aggressive package on the order of 0.6 plants. That is roughly a 10× gap from Conservative to Baseline and a further ~6–7× gap from Baseline to Aggressive at the same corridor point.

Taken together, Figures 8 and 9 show that the deployment translation produces two distinct outputs that must be read jointly. The first is the magnitude of net grid relief at the campus boundary. The second is the infrastructure scale needed to obtain a given displacement target at a given corridor distance. The two are not the same object, and they can move in opposite directions across packages.

#### 3.6.5. Interpretation and Deployment Implications

The deployment-scale figures sharpen how WtE-coupled cooling should be evaluated as infrastructure. The net grid relief in Figure 8 is economically meaningful only when paired with the corridor sizing in Figure 9. The resulting interpretation is a three-layer feasibility logic: (i) the amount of cooling electricity available to be displaced, (ii) the deliverable low-grade heat that can arrive over a corridor, and (iii) the WtE supply footprint required to close the gap. This distinction is especially important because AI DCs are typically sited around power interconnections, fiber routes, and developable land, whereas WtE facilities are sited around municipal waste basins, transfer logistics, and permitting constraints. The representative plant counts reported here should therefore be read as scale-mismatch indicators, not as literal local plant counts. Large implied counts signal that a concept may require a multi-node regional program, transfer-network integration, or a looser interpretation of the displacement target, rather than immediate physical co-location at the campus boundary.

First, grid relief is bounded by the baseline cooling-electricity block, which differs sharply across packages. The three $\Delta Grid$ values, 264.4 MW, 195.5 MW, and 116.9 MW, track the size of the mechanical-cooling segment in the Standalone bars. This directly limits the maximum benefit of coupling under the chosen boundary. The Aggressive package illustrates the key point: high baseline efficiency reduces the available displacement margin. A smaller $\Delta Grid$ is therefore not an unfavorable outcome in itself. It can reflect a campus that already purchases relatively little "support electricity" per unit IT load. In infrastructure terms, it implies that the coupled pathway should be justified more on capacity relief under binding interconnection constraints or on portfolio diversification, rather than on large incremental electricity displacement.

Second, the required WtE footprint is the decisive feasibility determinant in the Conservative package. At $L = 5$ km, the Conservative package implies roughly 40 plants to reach 25% displacement and roughly 80 plants to reach 50% displacement. Figure 9's full-coverage reference case (annotated as $N \approx 148$) sits on the same scale. Even before considering contracting and permitting constraints, this magnitude transforms the project concept. It is no longer a "single-node coupling" but a coordinated multi-plant program, likely spanning multiple municipal counterparties and waste-supply basins. The corridor remains short in absolute terms, but the implied supply footprint is large. In practice, feasibility would likely require either much larger plant units than the representative benchmark, tighter co-location, or an explicit reinterpretation of "coverage" toward partial displacement targets. The Conservative package is therefore best read as a lower-bound feasibility test that highlights when corridor losses and limited exportable heat make coupling infrastructure intensive.

Third, baseline and Aggressive packages support a corridor strategy with fundamentally different deployment modes. Baseline suggests that meaningful displacement targets can be achieved with a moderate fleet. At $L = 5$ km, approximately four plants support 25% displacement and about eight plants support 50% displacement. Those magnitudes are compatible with corridor planning at the metropolitan scale, where a small number of large WtE facilities can be within a few to tens of kilometers of a campus. Aggressive further compresses the implied footprint. At $L = 5$ km, the 50% curve sits near one plant, and the 25% curve is well below one plant in full-load-equivalent terms. That places the deployment mode in a "single anchor node" regime, where discrete siting becomes dominant and operational integration questions matter more than siting feasibility. The part-load indicator in Figure 8 (approximately 0.76 for $N = 3$) is consistent with this regime: the infrastructure can be installed in small integers, and the design question shifts toward matching heat availability, storage, and cooling dispatch rather than toward expanding the supply footprint.

Fourth, distance plays two different roles: a marginal sizing penalty in Baseline/Aggressive, and a binding corridor constraint in Conservative. Over 0–30 km, Conservative's plant requirements grow visibly, while Baseline rises only slightly and Aggressive is almost flat. This is the practical meaning of $\beta$ in Appendix A.4. It is not merely a modeling convenience. It governs whether the corridor behaves like a gradually increasing cost term or like a feasibility cliff. When $\beta$ is high, corridor extension translates into a multiplicative increase in infrastructure requirement, and the integration must be treated as a narrow corridor strategy. When $\beta$ is low, corridor length over tens of kilometers is not the binding constraint, so feasibility hinges on other elements such as waste supply contracting, dispatch priority, and local permitting.

Fifth, coverage targets are infrastructure commitments, not parameter tweaks. The nearly proportional scaling from 10% to 25% to 50% displacement within each package implies that increasing the target coverage roughly multiplies the required plant count. This matters because a 25% target in Baseline implies approximately four plants at 5 km, while moving to 50% implies approximately eight plants. The incremental "benefit" in grid relief is not free; it requires a discrete expansion of the WtE supply footprint or a tighter corridor. In Conservative, the same move shifts the project from dozens of plants to near a hundred at short distances. Coverage choice therefore belongs in early-stage infrastructure planning, not in late-stage engineering optimization.

Lastly, discrete siting and part-load operation introduce a second layer of operational constraints. When the required plant count is in the low single digits, rounding to integers can introduce the part-load operation of installed capacity. The Aggressive package illustrates this directly. Part-load does not need to reduce net benefit if the system is designed to absorb mismatch. However, it raises the relative importance of storage buffering and dispatch logic. In this regime, CTES/TES is no longer an optional enhancement. It becomes a natural complement for maintaining high displacement during grid-constrained hours while accommodating variability in WtE heat availability and campus cooling demand. The screening translation does not model these dynamics, but it provides quantitative bounds that are needed to size them.

## 4. Discussion and Implications: Thermal Management, Renewable Conversion, and Decision Metrics

### *4.1. Further Discussion of the Main Findings*

The coupled WtE–AIDC configuration is best interpreted as an infrastructure pathway for cooling-service procurement, where low-grade thermal output is converted into delivered cooling that displaces electricity-driven chilling at the campus boundary. In the reference illustration, a modern regulated WtE export stream of $Q_{drive,0} \approx 78.1\ \mathrm{MW_{th}}$ translates into $W_{avoid,net}(L) \approx 8.9\mathrm{e}^{-0.005\mathrm{L}}\ \mathrm{MW_e}$, which remains on the order of 7–9 $\mathrm{MW_e}$ over 0–40 km and yields a coverage threshold distance of $L^{*} \approx 21$ km for a 40 MW IT site with an 8 MW cooling-electricity target. Distance then becomes a first-order planning constraint rather than a convenience, because the same configuration exhibits a coverage threshold $L_{max} \approx 20.9$ km and a thermoeconomic viability frontier $L_{max} \approx 50.3$ km, beyond which attenuation and parasitics dominate the recovered cooling value. This corridor logic implies a non-linear feasibility map: when deliverability enters through $e^{-\beta L}$, incremental distance can have multiplicative sizing consequences, so corridor selection is a discrete infrastructure decision rather than a marginal engineering adjustment.

The utilization and feedstock sensitivities share a single regime structure governed by the coverage ratio $f(L)$: improvements in thermal supply raise $\Delta PUE_{elec}$ only in partial coverage, while gains plateau once $f(L) = 1$ and residual electricity is pinned by fixed overheads and parasitics. In this sense, COP values alone do not determine the system's infrastructure value, because the binding constraint is often dependent on whether sufficient deliverable cooling is available to move the campus from partial to full coverage at a feasible corridor length. The quality-adjusted efficiency results reinforce the same point using a different accounting lens: the incremental benefit does not rely on optimistic generation performance, but on grade matching that reallocates a fixed primary energy input toward outputs aligned with the campus' low-grade cooling demand.

Deployment-scale translation makes the decision objects explicit in units that planners can act on. For a 1 GW IT campus at $\mathrm{u} = 0.70$ and $\mathrm{L} = 5$ km, the coupled ledger yields a net grid relief of $\Delta Grid \approx 116.9$–264.4

MW across scenario packages, while the implied WtE footprint spans $N \approx$ 3–148 of the equivalent representative WtE plants under the representative plant convention, or approximately 0.6–40 full-load-equivalent plants at a 25% displacement target. These magnitudes separate benefit from feasibility: large $\Delta$grid tends to coincide with a large baseline cooling-electricity block, whereas implementability is governed by corridor deliverability and the institutional scale of the required WtE fleet, including discrete siting effects that can induce part-load operation even when displacement targets are met. This separation motivates the remainder of Section 4, which treats sustainability boundaries and attribution, operational flexibility extensions, metric interfaces for cost and reporting, corridor planning constraints, and limits to external validity as distinct layers of the same infrastructure decision problem.

*4.2. Sustainability Mechanisms and ESG Boundary*

The sustainability relevance of WtE–AIDC coupling follows two measurable mechanisms, provided that reporting remains explicit about system boundaries and counterfactuals. The first channel is waste diversion from landfills to WtE, which changes the methane pathway under a stated landfill management regime. The controlled operational object is the diverted quantity and composition of waste, $W_t$, and credible attribution requires pairing $W_t$ with a transparent counterfactual specification rather than asserting a single-point estimate of methane abatement. The second channel is electricity displacement for cooling, since thermal coverage reduces the share of cooling met by mechanical chillers at the campus boundary. This effect is directly observable through $\Delta PUE_{\text{elec}}$ under the stated utilization and corridor conditions, and through normalized electricity intensity $E_t^{grid}/K_t$ using the same compute-service denominator adopted for cost interfaces in next sections. Importantly, the results in Sections 3.4 and 3.5 impose sharp conditions on what can be claimed: efficiency and emission implications are regime-dependent, and should be reported as contingent on whether the system operates in partial coverage or full coverage and whether the corridor lies within the viability window implied by $L_{\max}$. Within these boundaries, a disciplined ESG template is straightforward: disclose $W_t$ and its counterfactual treatment path, disclose the grid-electricity intensity of computing and its dependence on $f(L)$, and disclose the boundary assumptions used to translate thermal coupling into avoided grid electricity. Across adjacent sustainability domains, recent studies have combined interpretable resilience and socioeconomic indicators, carbon-market forecasting, and process-level air- and water-pollution control to support transparent, scenario-specific decisions [43–48].

*4.3. Operations and Flexibility Extension*

The screening results are steady-state, yet the same boundary logic extends naturally to operational flexibility when cold or thermal energy storage (CTES/TES) is introduced. Storage primarily reshapes when thermal-derived cooling is delivered, not what the coupled system is at the reporting boundary. This distinction matters because the regime structure identified in Section 3 implies that storage is most valuable near the corridor frontier and in partial-coverage operation, where delivered heat is marginal and small shifts in effective coverage can meaningfully reduce residual mechanical chilling. In a storage-enabled configuration, absorption cooling can charge CTES during periods of higher heat availability or lower auxiliary burden, then discharge during grid-constrained or high-price hours to deepen cooling-electricity displacement without changing the accounting identities used to compute $E_t^{grid}$. The same coverage and corridor thresholds therefore provide first-order bounds for storage sizing and expected value, while detailed dispatch and reliability constraints can be layered on through rule-based control or MPC. The operational extension is best viewed as a feasibility-preserving add-on that increases the robustness of the coupling channel rather than as a different value proposition. A short-interval co-optimization between plant-side electricity production, exportable heat, storage buffering, and cooling dispatch is beyond the scope of the present screening paper; the current contribution is to make the hybrid service logic and reliability-preserving fallback explicit at corridor-planning level.

*4.4. Economic Interpretation of Methodology-Defined Metrics*

Using the decision metrics defined in Section 2, the corridor results from Section 3 can be expressed as an economic comparison between the baseline electricity-driven cooling configuration and the WtE-coupled alternative. The purpose of this layer is not to construct a full project-finance model, but to provide a boundary-consistent interface that links thermodynamic screening to planning-relevant cost interpretation.

In this framework, delivered cooling, net avoided electricity, and the coverage ratio $f(L,\rho)$ determine how much of the baseline cooling requirement remains grid-served. Under partial coverage, additional thermal availability continues to displace mechanical chilling and reduce electricity purchases. Once full coverage is reached, further thermal improvement no longer lowers cooling-related grid demand under the stated bookkeeping; distance and feedstock then matter mainly through parasitic burden, corridor feasibility, and the durability of the coupled configuration.

LCOC should therefore be interpreted as a comparison interface rather than a standalone cost verdict. Its value is that it preserves denominator consistency while translating the regime logic of Sections 3.2–3.5 into an economic object: favorable comparisons are most likely where corridor conditions sustain strong coverage with limited auxiliary penalty, while long-distance or low-quality cases compress that advantage and eventually move the system outside the admissible planning region [42].

### *4.5. Deployment Constraints and Corridor Planning*

The combined evidence supports interpreting WtE–AIDC integration as a corridor strategy rather than a universal design rule. The key planning objects are the distance thresholds that define regime shifts and viability. The coverage boundary $L_{max}$ marks where the system transitions from full coverage to partial coverage, and the thermoeconomic frontier $L_{max}$ marks where attenuation and parasitics overwhelm recovered cooling value. These thresholds convert thermodynamic logic into siting constraints that are actionable at the infrastructure planning stage. They also discipline the interpretation of PUE gains, because plateaus in $\Delta PUE_{elec}$ arise when coverage is already complete, not because the technology ceases to improve.

Deployment-scale translation adds a second constraint that is often decisive in practice: the implied infrastructure footprint required to reach a displacement target at a given corridor distance. The per-GW results in Section 3.6 show that meaningful grid relief can coexist with widely different WtE supply requirements, so feasibility depends jointly on corridor deliverability and institutional scale. This coupling between physics and footprint provides a disciplined basis for phased expansion. Campus growth that increases effective separation to new IT halls can trigger discrete moves from full coverage to partial coverage, or from viability to non-viability, so expansions that cross $L_{\max}$ or approach $L_{\max}$ should be evaluated as requiring an additional thermal node, a different cooling architecture, or a revised contract structure rather than being priced as marginal capacity on the original curve. The same corridor framing aligns with location-robust planning perspectives for multi-campus AI infrastructure, where heterogeneous siting constraints and energy conditions are first-order determinants of outcomes. Future project-calibrated implementations could extend the present screening logic with explainable data valuation, digital-twin and intelligent-industrial representations, and structured knowledge graphs for integrating heterogeneous plant, corridor, and campus evidence [51–54]. Document-grounded retrieval, routing, and reranking could further support auditable updates from engineering, regulatory, and financial sources [55,56]. If LLM-based interfaces are introduced, prompt formulation should be standardized and evaluated, while system-aware serving design should be considered separately from facility-level energy accounting [57,58].

## 5. Conclusions

This paper developed a boundary-consistent screening framework for evaluating WtE-coupled AI DC cooling as an infrastructure problem rather than as an equipment-level design exercise. The core objective was to determine when low-grade WtE heat can be translated into delivered cooling, net grid-electricity relief, and corridor-feasible deployment under explicit distance and parasitic constraints.

Three findings emerge. First, the reference-case energy-service ledger shows that a representative regulated WtE plant can generate decision-relevant cooling value once corridor attenuation, absorption conversion, and auxiliary electricity are all kept inside the same accounting boundary. In the baseline reference case, a 1500 t/day plant with LHV of 10 MJ/kg provides about 78.1 MWth of exportable driving heat and yields about 8.0 MWe of net avoided cooling electricity at 20 km, remaining positive at 40 km. This makes corridor separation a quantitative planning variable rather than a qualitative siting preference. Second, the coupled system is governed by operating regimes rather than by a single efficiency score. The revised results show a clear distinction among full coverage, partial but viable operation, and non-viable corridor conditions. Under the baseline package, the full-coverage threshold is about 20.9 km, while the stricter quality-adjusted threshold is about 22.9 km; under conservative and aggressive assumptions, these corridor bands contract or widen materially. The result is a siting-ready feasibility map rather than a single headline performance number. Third, deployment-scale translation separates benefit from implementability. At campus scale, heat-driven cooling can produce material grid relief, but the implied WtE footprint varies sharply across scenario packages and displacement targets. The practical implication is that coupling may be attractive in energy terms while still being difficult in institutional or spatial terms.

In addition, a project-calibrated extension would need to integrate corridor thermodynamics with regional MSW-basin availability, transfer logistics, permitting, and DC siting around fiber/power nodes. Those spatial-compatibility questions are intentionally separated from the present first-pass screening framework.

To sum up, these results support a narrower but more actionable conclusion than a generic claim about WtE or AI infrastructure. WtE-coupled cooling is best treated as a conditional corridor strategy for relieving cooling-related electricity demand where proximity, deliverability, and auxiliary burdens remain favorable. The contribution of the paper is therefore a planning framework that identifies when such pairings remain plausible before detailed engineering begins. Detailed turbine-cycle modeling, routed thermo-hydraulic design, storage dispatch, permitting, and GIS-based compatibility with waste basins, power nodes, and fiber corridors are left to project-calibrated follow-on work. Within that boundary, the present framework provides a compact and auditable basis for screening WtE–AIDC pairings under uncertainty.

**Author Contributions:** Conceptualization, Q.H. and C.Q., and W.Z.; methodology, Q.H. and C.Q.; formal analysis, Q.H. and C.Q.; investigation, Q.H. and W.Z.; data curation, Q.H. and C.Q.; writing—original draft preparation, Q.H.; writing—review and editing, Q.H. and C.Q., and W.Z.; visualization, Q.H.; supervision, C.Q., and W.Z.; project administration, Q.H. and W.Z. All authors have read and agreed to the published version of the manuscript.

**Funding:** This research received no external funding. The APC was funded by the authors.

**Data Availability Statement:** The code, computational materials, and supplementary Appendix A.2–A.4 used to support the findings of this study are available at GitHub: https://github.com/leondardoq/Planning-WtE-AI-DC-through-Grade-Matched-Cooling-and-Corridor-Screening (accessed on 19 April 2026). No new proprietary dataset was created. The screening inputs and parameter values needed to reproduce the reported results are provided in the manuscript, especially in Tables 2–3 and Appendix A.

**Conflicts of Interest:** The authors declare no conflict of interest.

## Appendix A. Proof of Proposition 1 (A Sufficient Condition for Thermoeconomic Superiority)

This appendix provides a formal derivation of Proposition 1 stated in Section 3.4. Throughout, we compare two architectures under a common system boundary and a common primary energy input from waste.

### *Appendix A.1. Setup and Accounting Conventions*

Let the primary energy input rate from waste be $\dot{E}_{in} = \dot{m}_w LHV$. Let $\eta_c$ denote effective conversion/utilization within the WtE boundary (combustion/boiler availability), so the primary exergy/available energy entering the conversion chain is approximated as $Ex_{in} \equiv \eta_c \dot{E}_{in}$. This is a standard screening approximation for comparative statics; the proof below only requires that $Ex_{in}$ is identical across configurations.

We define two configurations:

- Standalone configuration (S): WtE produces electricity $\dot{W}_e$. Recoverable heat is rejected (no credited useful output within the considered service set). DC cooling is provided mechanically using electricity.
- Coupled configuration (C): The same WtE produces $\dot{W}_e$. A portion of recoverable heat is delivered and converted into cooling. This requires additional parasitic work $\dot{W}_{par}$ (pumps/controls/auxiliaries for thermal delivery and absorption). We credit the cooling service by its exergy content.

Appendix A.1.1. Credited Useful Outputs

We credit two services as useful outputs within the coupled boundary:

- Electricity available for externally useful purposes (including supplying computing loads).
- Cooling service delivered to the DC, credited by its exergy content.

Let $T_0$ be ambient temperature and $T_c$ be the chilled-water (or evaporator) temperature. The exergy factor of a cooling effect is $\phi_c \equiv \frac{T_0}{T_c} - 1 > 0$, so the exergy of a cooling flow $\dot{Q}_{cool}$ is $Ex_{cool} \equiv \phi_c \dot{Q}_{cool}$.

Appendix A.1.2. Net Useful Exergy Output

A key point is that parasitic work $\dot{W}_{par}$ is not "free": it consumes high-grade electricity and should be treated as a reduction in net useful output within the same boundary. Hence, we define net useful exergy output as

Standalone: $Ex_{out,net}^{(S)} \equiv \dot{W}_e$

Coupled: $Ex_{out,net}^{(C)} \equiv \dot{W}_e - \dot{W}_{par} + \phi_c \dot{Q}_{cool}$.

This net-output definition is conservative and audit-friendly: it credits cooling only by its exergy content and explicitly debits parasitic high-grade work.

*Appendix A.2. Proposition and Proof*

Appendix A.2.1. Proposition 1 (Sufficient Condition)

Holding the primary input $Ex_{in}$ fixed, a sufficient condition for the coupled architecture to weakly dominate the standalone architecture in systemic exergy utilization is

$$\phi_c \dot{Q}_{cool} \geq \dot{W}_{par} \quad \text{(A1)}$$

Appendix A.2.2. Proof

Given identical $Ex_{in}$ across configurations, the dominance in exergy efficiency is equivalent to dominance in net useful exergy output, because

$$\eta_{ex,net}^{(k)} \equiv \frac{Ex_{out,net}^{(k)}}{Ex_{in}}, k \in \{S, C\}$$

Thus $\eta_{ex,net}^{(C)} \geq \eta_{ex,net}^{(S)}$ if and only if $Ex_{out,net}^{(C)} \geq Ex_{out,net}^{(S)}$.

Compute the difference:

$$Ex_{out,net}^{(C)} - Ex_{out,net}^{(S)} = \left(\dot{W}_e - \dot{W}_{par} + \phi_c \dot{Q}_{cool}\right) - \dot{W}_e = \phi_c \dot{Q}_{cool} - \dot{W}_{par} \quad \text{(A2)}$$

Therefore, if $\phi_c \dot{Q}_{cool} - \dot{W}_{par} \geq 0$, then $Ex_{out,net}^{(C)} \geq Ex_{out,net}^{(S)}$, implying $\eta_{ex,net}^{(C)} \geq \eta_{ex,net}^{(S)}$.

This proves (A1) is sufficient.

*Appendix A.3. Interpreting the Condition as a Grade-Matching Criterion*

Condition (A1) can be read as a grade-matching inequality: the quality-adjusted benefit from converting low-grade heat into cooling must exceed the high-grade work required to deliver and realize that conversion.

Using the model in Section 3, $\dot{Q}_{cool} = COP_{abs}\,\eta_{tr}(L)\,\dot{Q}_h$, the sufficient condition becomes

$$\phi_c\, COP_{abs}\,\eta_{tr}(L)\,\dot{Q}_h \geq \dot{W}_{par} \quad \text{(A3)}$$

Because $\eta_{tr}(L)$ is decreasing in distance, (A3) immediately implies the existence of a distance-dependent viability region for thermoeconomic dominance.

*Appendix A.4. Remarks on Strength and Scope (Why "Sufficient"?)*

The condition is sufficient, not necessary: the coupled system may still be attractive under alternative objective functions (e.g., grid-capacity relief, carbon intensity reduction, or electric OpEx savings) even when (A1) fails.

The condition relies only on (i) consistent system boundaries, (ii) a conservative exergy credit for cooling, and (iii) the explicit debiting of parasitic high-grade work, precisely the elements that prevent metric-gaming critiques.